 \documentclass[preprint2]{aastex}

\newcommand{{\msun}}{M$_\sun$}

\shorttitle{Intermediate-Age and Old stellar population models}
\shortauthors{V\'azquez \& Leitherer}

\begin{document}


\title{Optimization of Starburst99 for  \\
    Intermediate-Age and Old Stellar Populations}


\author{Gerardo A. V\'azquez and Claus Leitherer}
\affil{Space Telescope Science Institute, 3700 San Martin Drive, Baltimore, 
Maryland 21218}
\email{vazquez@stsci.edu \& leitherer@stsci.edu}


\begin{abstract}

We have incorporated the latest release of the Padova models into the evolutionary synthesis code Starburst99. The Padova tracks were extended to include the full asymptotic giant branch (AGB) evolution until the final thermal pulse over the mass range 0.9 to 5~{\msun}. With this addition, Starburst99 accounts for all stellar phases that contribute to the integrated light of a stellar population with arbitrary age from the extreme ultraviolet to the near-infrared. AGB stars are important for ages between 0.1 and 2~Gyr, with their contribution increasing at longer wavelengths. We investigate similarities and differences between the model predictions by the Geneva and the Padova tracks. The differences are particularly pronounced at ages $> 1$~Gyr, when incompleteness sets in for the Geneva models. We also perform detailed comparisons with the predictions of other major synthesis codes and found excellent agreement. Our synthesized optical colors are compared to observations of old, intermediate-age, and young populations. Excellent agreement is found for the old globular cluster system of NGC 5128 and for old and intermediate-age clusters in NGC 4038/39. In contrast, the models fail for red supergiant dominated populations with sub-solar abundances. This failure can be traced back to incorrect red supergiant parameters in the stellar evolutionary tracks. Our models and the synthesis code are publicly available as version 5.0 of Starburst99 at http://www.stsci.edu/science/starburst99/.

\end{abstract}




\section{Introduction}

Starburst99 is a multi-purpose evolutionary synthesis code for modeling various properties of stellar populations. The code was first used by Leitherer \& Heckman (1995) for compiling a set of observable parameters of evolving young stellar populations. The Leitherer \& Heckman models relied on an earlier release of stellar evolutionary tracks by Maeder (1990). These tracks were superseded by a new set of models from the Geneva group (Meynet et al. 1994 and references therein) so that an update to the original synthesis code seemed appropriate. This was done by Leitherer et al. (1999), who implemented the latest set of the Geneva tracks, added the capability of performing isochrone synthesis, and released a web-based interface to offer users interactive model computations at www.stsci.edu/science/starburst99/. Several important additional aspects were addressed over the years: Leitherer, Robert, \& Drissen (1992) discussed the mechanical energy release of evolving stellar populations; Leitherer, Robert, \& Heckman (1995) studied the satellite-ultraviolet (UV) spectra at high spectral resolution; de Mello et al. (1998) extended the latter work to less massive B stars; Leitherer et al. (2001) added an empirical UV library of metal-poor stars; Robert et al. (2002) included O-star spectra obtained with the FUSE satellite; most recently, Smith, Norris, \& Crowther (2002) implemented the latest generation of blanketed non-LTE model atmospheres. 

Starburst99, like any other evolutionary synthesis code, can in principle be applied to stellar populations having arbitrary properties. In practice, the useful parameter space that can be reliably predicted is determined by the set of stellar atmospheres, spectral templates, and evolutionary tracks. Starburst99 relies on the most advanced atmospheres and spectra for hot, massive stars (Lejeune, Cuisinier, \& Busser 1997; 1998; Hillier \& Miller 1998; Schmutz 1998; Pauldrach, Hoffmann, \& Lennon 2001) and their coupling to the Geneva evolutionary models (Schmutz, Leitherer, \& Gruenwald 1992). This property makes Starburst99 and similar codes (e.g., Schaerer \& Vacca 1998, Cervi\~no, Mass-Hesse, \& Kunth 2002 and references therein) uniquely qualified for modeling populations containing hot massive stars, in particular Wolf-Rayet (W-R) stars. On the downside, Starburst99 is not competitive when it comes to stellar populations whose properties are dominated by intermediate-age and old stars. This regime is better described by the stellar evolution models of the Padova group (Fagotto et al. 1994b and references therein) whose input physics pays particular attention to intermediate- and low-mass stars. Furthermore, the Padova models offer the possibility to extend the post-main-sequence evolution of intermediate-mass stars until and including thermally pulsing asymptotic giant branch (TIP-AGB) stars. These properties make the Padova tracks the models of choice for synthesis codes studying populations with ages above $\sim 10^8$ years. The most important synthesis codes concentrating on this age range are, e.g., those of Worthey (1994), Vazdekis et al. (1996), Fioc \& Rocca-Volmerange (1997), and Bruzual \& Charlot (2003).

Over the years, it has become customary to model young and old stellar populations with the codes that were optimized for the parameter space of interest and perform a comparison between the codes in the overlap region at intermediate age where either model is supposed to be applicable (e.g. Kewley et al. 2001; Whitmore \& Zhang 2002; Lamers et al. 2002). Very often, however, such a comparison is complicated by the fact that additional ingredients (such as model atmospheres) vary between different codes. Therefore our group has expanded the capabilities of Starburst99 to allow for a self-consistent treatment of stellar phases starting with and past the early asymptotic giant branch (EAGB) phase. In this paper we describe the upgrade done in Starburst99 and how it affects the model predictions. Since Starburst99 now offers models based on both the Geneva and the Padova tracks, relative comparisons can be done in a reliable way.

An additional major goal of this paper is a critical evaluation of the reliability of our evolutionary synthesis models when red supergiants (RSG) dominate. Stellar evolution models are notoriously uncertain for RSGs, in particular at low metallicity (Massey \& Olsen 2003). Therefore, comparisons of synthesis models with extragalactic systems containing massive cool stars are badly needed to provide empirical guidance for stellar evolution models.

This paper is organized as follows. First, we review similarities and differences between the main astrophysical ingredients in both the Geneva and Padova evolution models (Section~2). In Section~3 we investigate how those ingredients affect the main stellar predictables from both sets in the Hertzsprung-Russell diagram (HRD). A technical discussion of implementation issues in Starburst99 is in Section~4. Our main focus during the implementation was the treatment of the AGB stars. We discuss our approach in Section~5. A comprehensive comparison between the predicted properties of a standard population using either the Geneva or the Padova tracks in Starburst99 is performed in Section~6. In Section~7 we extend this comparison to other synthesis codes. Comparisons of old, intermediate-age, and young populations are performed in Section~8. Finally, in Section~9 we provide the conclusions, as well as guidelines for using the new models in Starburst99.

\section{Main physical ingredients in the Padova and Geneva tracks}

In this section we will compare the major physical ingredients in the two sets of evolutionary models. This assessment will help us identify the root causes of the differences in the predicted stellar properties of the two model sets, and eventually in the differences of the population properties synthesized by Starburst99. Throughout this paper we will use the ``high mass loss'' tracks of Geneva, as opposed to the ``standard'' tracks. The former are the recommended set whose properties have been fine-tuned to match the observed properties of massive stars (Meynet et al. 1994; but see Massey 2003 for an opposing point of view). The relevant literature sources for the Geneva models are Schaller et al. (1992), Schaerer et al. (1993a,b), Charbonnel et al. (1993), and Meynet et al. (1994). 

The Padova models used here are those for intermediate and low-mass stars in the range of 0.15 - 7~{\msun} from Girardi et al. (2000) and the tracks for massive stars in the range of 9 - 120~{\msun} from the original set of Padova models discussed in Bressan et al (1993), Fagotto et al. (1994a,b), and Girardi et al (2000). Throughout this work we will refer to this composite set of tracks as ``Padova models''.

We will focus our discussion in this section on the adopted chemical composition, the nuclear opacities and reaction rates, the convective energy transport, the mass-loss rates, and the boundary conditions at the stellar surface.

\subsection{Chemical composition}

Both sets of tracks use the solar-abundance tables of Anders \& Grevesse (1989) as a baseline, with updated values for carbon, nitrogen, and iron from Grevesse \& Anders (1991), Grevesse et al. (1991), Hannaford et al. (1992), and Grevesse \& Noels (1993). Note that these abundances are still on the older solar oxygen abundance scale with $\log$(O/H) + 12 = 8.9. More recent work by Allende Prieto, Lambert, \& Asplund (2002) has led to a downward revision of the solar oxygen abundance. The currently preferred value is $\log$(O/H) + 12 = 8.69.

Both the Padova and the Geneva models adopt the Anders \& Grevesse values for the zero-age-main-sequence (ZAMS) abundances of all elements other than He. (We will discuss the helium abundances further below.) In Table~1 we summarize these values for the most important elements up to iron. In the following we will adopt these conventions and nomenclature: abundances of individual elements, when the elements are denoted by their chemical symbols, are always by {\em number} on a scale where the hydrogen abundance equals 12. Following the nomenclature used in stellar evolution, we will use {\em X}, {\em Y}, and {\em Z} for the {\em mass} fractions of hydrogen, helium, and all elements heavier than helium, respectively. We will avoid the incorrect use of ``metallicity'' to denote {\em Z}. Metallicity should be restricted to the logarithmic number abundance of metals relative to the sun. 

The heavy-element composition for non-solar abundance models in both the Geneva and the Padova tracks was derived from the solar-abundance models by applying an appropriate scaling of the heavy-element abundances and by keeping the metal-abundance ratios fixed. Small adjustments to the metal-abundance ratios were required in order to be consistent with the heavy-element ratios used for the opacity calculations (Schaller et al. 1992). Both Padova and Geneva published five complete sets of evolutionary tracks with chemical compositions covering a wide range of astrophysical applications. Three of the five sets have identical metal abundances in Geneva and Padova: $Z = 0.02$ (solar), 0.008 (40\% solar), and 0.004 (20\% solar). In addition, Padova published super-solar models with $Z = 0.05$ (2.5 solar) for the full mass range and with $Z = 0.03$ (1.5 solar; Girardi et al. 2000) with updated opacities for intermediate- and low-mass stars. Our composite set of Padova tracks uses a merged version of the $Z = 0.05$ and 0.03 tracks for massive and intermediate$+$low-mass stars, respectively. For comparison, Geneva published models with $Z = 0.04$ (twice solar). Padova and Geneva also published extremely metal-poor tracks with $Z = 0.0004$ (2\% solar) and 0.001 (5\% solar), respectively. The differences in the heavy-element abundances should be kept in mind when making comparisons between different tracks. 

The helium abundances follow from the solar and primordial helium abundances, respectively, and the adopted ${\Delta Y}/{\Delta Z}$ relation. The Padova models use a primordial helium abundance of 0.24 (note that Girardi et al. 2000 use 0.23) and ${\Delta Y}/{\Delta Z} = 2.5$. The Geneva group prefers a slightly steeper ${\Delta Y}/{\Delta Z} = 3$ for sub-solar and solar models, whereas for super-solar models they use ${\Delta Y}/{\Delta Z} = 2.5$, in agreement with Padova. Although the helium abundance is important in controlling the opacity and determining the core temperature profile, the effects of varying $Y$ become dominant only at high metal abundance (Yi 2003).

A compilation of the ZAMS abundances of H, He, C, N, and O for the five metal abundances is in Table 2. In summary, we find very similar chemical abundances for both the Geneva and Padova models. Except for the very metal-rich and -poor model sets, different abundance profiles cannot be the reason for differences between the tracks.

\subsection{Nuclear opacities and reaction rates}

The opacities in both sets of evolutionary models are from the widely used Livermore group (e.g., Iglesias, Rogers, \& Wilson 1992 and references therein).  The reaction rates adopted by the Padova and Geneva models are different. The largest disagreement is for the rate of $^{12}$C($\alpha$, $\gamma$)$^{16}$O in the CN cycle. The Geneva group developed its models with the rates calculated by Caughlan et al. (1985), whereas the Padova group used those of Caughlan \& Fowler (1988). The latter calculations result in a lower reaction rate.

The effects of varying the rate of this reaction on evolutionary models have been known since the early study by Iben (1972). The larger cross section leads to a higher conversion rate of carbon into oxygen and to more pronounced loops extending to the blue part of the HRD before rapid core contraction and envelope expansion set in, and the evolution proceeds back to the red. Overall, the core-He-burning lifetime is increased by several percent. In practice, however, these effects are difficult to separate from those of mass loss and convective overshoot (Chiosi 1998).

\subsection{Convective energy transport}

A comprehensive discussion of the importance of the convective energy transport for stellar evolution models and synthetic stellar populations was given by Yi (2003). While Yi's work is tailored towards intermediate- and low-mass stars, many of the arguments are also qualitatively correct when applied to massive stars. In particular, the efficiency of the convective energy transport has profound consequences on the evolution of massive stars. Massive stars have convective cores and radiative envelopes, except for the W-R phase. Increasing the size of the convective core, e.g., by convective overshooting, will increase the rate of the energy transport since convection is more effective than radiation. Therefore the surface temperature will critically depend on these assumptions.

The prescription for the convective overshooting in the Padova tracks is as follows: low-mass stars have a core overshoot of $d/H_{\rm{P}} = 0.25$, where $d$ is the overshoot distance and $H_{\rm{P}}$ is the pressure scale height, whereas more massive stars have a value of 0.5. In contrast, this parameter is set to 0.2 in the Geneva models, as suggested by an empirical calibration over the mass range from 1.25 to 25~{\msun} (Schaller et al. 1992). Fagotto et al. (1994a) demonstrated that the Padova and Geneva overshoot parameters are in fact consistent, as the algorithm for calculating the overshoot distance is different in the two sets of evolution models.

The mixing length parameter $\alpha = l/H_{\rm{p}}$, with $l$ being the mixing length, was derived by the Geneva group from an empirical calibration of their models via the observed location of the red giant branch in 75 star clusters (Schaller et al. 1992). $\alpha = 1.6 \pm 0.1$ was found and adopted. The Padova models use a very similar mixing length parameter of $\alpha = 1.63$ (Bressan et al. 1993).

Both sets of models make rather similar assumptions on the convective energy transport. However, neither set considers the effects of rotation. Rotation has been recognized to be a major driver of stellar evolution, in particular for massive stars (Meynet \& Maeder 2003, and references therein). Systematic models suitable for implementation into synthesis models are under construction. When such models will become available, their post-main-sequence evolution may differ from that in the current Padova and Geneva models.

\subsection{Mass-loss rates}

Stellar mass loss is a fundamental parameter for the evolution of massive stars. Both Geneva and Padova use the parameterization of de Jager et al. (1988) who compiled empirical mass-loss rates across the entire upper HRD. Deviating from this prescription, the Geneva group modified the prescription for the mass-loss rates of W-R stars and their evolutionary progenitors. All massive stars with radiatively driven winds (i.e., OB and WNL stars) are assumed to have twice the de Jager mass-loss rates. The motivation for this deviation from de Jager's relation is purely empirical: the resulting W-R parameters in the HRD are in better agreement with the observations. The adjustment of the mass-loss rate is simply a proxy for other effects not accounted for, such as rotation and convection. The rates of WNE and WC stars were calculated with the prescription of Langer (1989), who showed that these stars tightly follow the mass-loss rate vs. mass relation $\dot M = (0.6-1.00) \times 10^7(M/M_\sun)^{2.5}$ [{\msun} yr$^{-1}$]. For the Padova tracks, the W-R phase sets in for a hydrogen surface abundance below 30\% (Bressan et al. 1993), whereas for the Geneva tracks this limit is 40\% (Schaller et al. 1992), in both cases the mass-loss rate from Langer's relation is used.

The mass-loss rates of stars with non-solar chemical composition were scaled with the relation $\dot M \propto Z^{0.5}$ (Kudritzki, Pauldrach, \& Puls 1987). This scaling was not applied in the W-R phase for the Geneva models. In contrast, the Padova tracks include this scaling of the mass-loss rates in all stellar phases.

Neither set of evolutionary models includes mass loss for low- and intermediate-mass stars, as the rates are too low to be significant. When we added the thermally pulsing AGB stars, we adopted the Reimers relation (Reimers 1975) for all evolved stars in the red part of the HRD with masses below 12~{\msun} outside the AGB phase.

\subsection{Boundary conditions at the stellar surface}

The most significant difference between the Padova and the Geneva tracks occurs in the W-R phase, when the definition of effective temperature ($T_{\rm{eff}}$) becomes ambiguous because of the highly extended atmospheres. The challenge is to define the radius and $T_{\rm{eff}}$ in the atmosphere. The calculation depends on the choice for the opacity in the wind of the W-R star. Langer (1989) and Maeder (1990) proposed to calculate the opacity from the continuity equation assuming a standard velocity law $V(r) \propto (1 - R/r)^{\beta}$, where {\em R} is the radius of the star and r is the distance. The simplifying assumption of pure electron scattering opacity ($\sigma_{\rm{e}}$) is made. The radius refers to the location where the optical depth $\tau(R_{\rm{eff}}) = 2/3$, and $T_{\rm{eff}}$ is determined from the Stefan-Boltzmann law at $R_{\rm{eff}}$.

In the case of the Geneva tracks, Schaller et al. (1992) followed Langer's (1989) suggestion, together with the Castor, Abbott, and Klein theory (Castor et al. 1975; Pauldrach et al. 1986; Kudritzki et al. 1989) to calculate the opacity. They adopted a mean weighted opacity $\kappa = \sigma_{\rm{e}}(1+FM)$ obtained from the radiation-driven wind theory, where {\em FM} is the so-called force-multiplier, which gives the contribution from absorption lines. This parameter is calculated from the radiation-hydrodynamics in the atmosphere. The final $R_{\rm{eff}}$ and $T_{\rm{eff}}$ are calculated by adding the line opacity to the pure electron scattering opacity. This correction for $T_{\rm{eff}}$ shifts the stellar positions to the red part of the HRD. Effective temperatures obtained with this correction cover the domain of the observed WNL, WNE, and WC stars shown as by Maeder (1990) and Schaerer \& Maeder (1992).

In the Padova tracks, there is no correction for the temperature decrease due to stellar winds. The boundary conditions simply follow from the optical depth calculated with the opacities of Rogers \& Iglesias (1992). Therefore, no outflow of material is allowed for, and all stars follow the same definition of $T_{\rm{eff}}$. As a result, W-R stars in the Padova models have higher temperatures than those in the Geneva models.

\section{Physical differences and their effects on the stellar evolution}

In this section we will analyze the effects of the different physical input parameters on the main evolutionary properties of individual stars in the HRD. First, we will discuss the principal parameters luminosity {\em L} and $T_{\rm{eff}}$ as a function of time. Then we will describe the differences in the surface abundances. We have chosen those three sets of tracks whose abundances are identical in the Padova and Geneva models ($Z = 0.004, 0.008, 0.02$). Both sets of tracks predict very similar stellar parameters (except in the W-R phase), with the solar metallicity tracks from the Padova group producing somewhat older ages than those of the Geneva group ($\sim 10\%$ and $\sim 4\%$ for intermediate and massive stars, respectively).

On the ZAMS, {\em L} and $T_{\rm{eff}}$ are very similar for most stellar masses in both model sets. Padova models are slightly hotter by $\sim 0.03$ dex and more luminous by $\sim 0.02$ dex than Geneva models. The similarity of the values is due to the same source of the opacities and abundances, and the similar treatment of overshooting. As discussed by Charlot, Worthey, \& Bressan (1996) and Yi (2003), the major effect is the overshooting, yet the opacities and the element abundances play an important role as well. Bruzual \& Charlot (2003) noted the 50-200 K higher temperature of the red giant branch (RGB) in the models of Girardi et al. (2000) in comparison with the original Padova set. Consequently, the Girardi et al. tracks produce bluer colors than those from the original Padova set in the Bruzual \& Charlot models after the appropriate color transformation is applied.

The evolution of {\em L} and $T_{\rm{eff}}$ as a function of time is very similar, except for the W-R phase where the models from Geneva are more luminous and cooler than the models from Padova. This behavior is expected due to the different boundary conditions on the stellar surface in the Geneva models. The ultimate astrophysical reason for the difference is not so much a real difference in $T_{\rm{eff}}$ but rather a different definition of temperature. $T_{\rm{eff}}$ in the Padova models refers to deeper atmospheric layers, whereas the Geneva definition samples the outer atmospheric region where the photons can escape (Schmutz, Leitherer, \& Gruenwald 1992). 

In the case of intermediate-mass stars, we note that both sets of models do not take into account the tip of the asymptotic giant branch (TIP-AGB). In order to minimize confusion, we avoid calling the end of the AGB phase the first thermal pulse in the AGB phase (TP-AGB) as opposed to the convention often used in other work. The behavior of $T_{\rm{eff}}$ is very similar in Geneva and Padova except when the horizontal branch (HB) is modeled. The Padova tracks have a different treatment for this phase. As a result, the temperature becomes higher than in the Geneva tracks. The behavior of these parameters in the AGB phase will change when we introduce the TIP-AGB for the Padova tracks as discussed in Section~5.

For the low-mass regime, the set of tracks from the Geneva group does not include the EAGB phase for stars with $0.8 \le M/M_\sun \le 1.7$, whereas the Padova tracks include the evolution up to the EAGB phase for stars $M \ge 0.6$~{\msun}. Starburst99 does not use the supplemental release of low-mass tracks by the Geneva group (Charbonnel et al. 1996). This release extends the early evolution of stars in the mass range $0.8 \le M/M_\sun \le 1.7$ to the EAGB phase. As a result, the Starburst99 implementation of the Geneva models in this mass range lacks complete coverage of even the RGB phase. Figure~1 shows tracks for the low-mass regime emphasizing the different phases covered by each group. The 1.7~{\msun} track is the transition between stars with and without a degenerate helium core in the Geneva models and are represented with a solid line and crosses, respectively, in Fig.~1. The track without a degenerate core has a clump of points which represent the counterpart of the EAGB phase in the Geneva tracks. This transition has been discussed by Girardi \& Bertelli (1998). The additional evolutionary phase in the extended 1.7 {\msun} Geneva track (and the tracks corresponding to lower masses) will produce noticeable features in the evolution of colors.

We also compared the evolution of He, C, N, and O as a function of time. For massive stars, the values reached at the beginning of and during the W-R phase are very similar at the three abundances. The final value of the oxygen abundance is $\sim 0.3$ dex higher in the Geneva tracks than in the Padova tracks (e.g., at 60 and 40~{\msun}). The carbon abundance is very similar in both groups, with a difference of $\sim 0.1$ dex at 60 and 40~{\msun}. For the same masses, the nitrogen abundance behaves similarly, except for models at $Z = 0.008$ where the Padova 40~{\msun} model has a much higher value than the Geneva model. This is the result of different limiting masses for the formation of W-R stars. These differences do not significantly affect the integrated parameters of synthetic populations. However, they do affect the census of W-R subtypes, which are defined by the surface abundance of these elements.

For intermediate stellar masses, we found abundance differences of $< 0.01$ dex in boths sets of tracks.

\section{Implementation of the Padova tracks in Starburst99}

We implemented the publicly available set of Padova models into Starburst99, closely following the scheme we used for the Geneva tracks. The models are distributed as tables with the pertinent stellar parameter given as a function of time for each stellar mass. The main differences between the Padova and the Geneva models, in addition to the previously identified W-R parameters, are the larger mass range and the more detailed coverage of late evolutionary phases of low- and intermediate-mass stars in the Padova tracks.

Two sets of Padova models were prepared for implementation. One set has the stellar evolutionary phases as specified in the Padova tables, and the second includes additional entries for thermally pulsing AGB stars. Apart from these additional entries, the two sets are identical, and both are available to users of Starburst99. We will discuss the specifics of the AGB phase in the subsequent section.

As we did in the Geneva tracks, we selected the initial mass $M_{\rm{i}}$, stellar age {\em T}, current mass $M_{\rm{c}}$, $\log L$, $\log T_{\rm{eff}}$, and the surface abundances of H, He, C, N, and O by mass fraction from the Padova tables. The original tables list $\dot M$ only for massive stars with masses $\geq 9$~{\msun}. The missing entries were compiled as follows. For the AGB phase, we used the values of Vassiliadis \& Wood (1994). For stars off the main-sequence with missing table entries, we applied the Reimers (1975) formula. Finally, we adopted low place holder values of $10^{-15}$~{\msun}~yr$^{-1}$ for less massive main-sequence stars. This choice is purely driven by computational reasons. Main-sequence stars with masses below 9~{\msun} are unimportant for the mass return of a population and, in addition, their mass-loss rates are poorly known. We performed these updates of the mass-loss rates only in the set of Padova tracks to which we added the TIP-AGB stars. The original Padova release was left unchanged. Note the absence of the core temperatures and radii, which are used in the Geneva tracks to distinguish wind-corrected W-R atmospheres from their uncorrected counterparts. We set the core temperatures and radii equal to $T_{\rm{eff}}$ and $R_{\rm{eff}}$, respectively, in our implementation of the Padova tracks. Furthermore, we assumed the same mass limits for the formation of W-R stars in the Padova models as in the Geneva models. The lowest stellar mass evolving into W-R stars depends on the metal abundance, as suggested by observations (Massey 2003). The Geneva values are $M_{\rm{i}} = 21$, 25, 35, 42, and 61~{\msun} for twice solar, solar, 40\%, 20\%, and 5\% solar composition, respectively (Maeder \& Meynet 1994). We adopted the same values for the Padova tracks with chemical composition 2.5 solar, solar, 40\%, 20\%, and 2\% solar, respectively.

A successful interpolation technique requires identification of the strategic evolutionary phases along the tracks. These are: the ZAMS point, the first temperature minimum prior to core-hydrogen exhaustion, the terminal-age main-sequence (TAMS), the epoch of core-hydrogen exhaustion, the supernova phase, the beginning of the RGB phase and the epoch of its luminosity maximum in the HRD, the onset of the horizontal branch evolution and its maximum luminosity, the epoch of core-helium exhaustion, and the beginning of the EAGB phase. During the interpolation, special attention is paid to these phases to ensure no unphysical result is obtained for those initial masses which are not explicitly specified in the distributed Padova tables.

After linking the Padova tables to Starburst99, they are treated identically to the Geneva tracks (see Leitherer et al. 1999 for details on the computational procedure). Either a traditional mass interpolation or an isochrone interpolation in time is performed. The latter technique is particular suitable for later evolutionary phases which would otherwise produce discontinuities in the synthesized quantities. Extensive test were performed to ensure the interpolation produces sensible results. We are confident that the implementation of the Padova models into Starburst99 with and without the AGB phases is as reliable as that of the Geneva tracks.

In addition to allowing for the AGB phases, the Padova tracks extend the valid mass range in Starburst99 from the previous 0.8~{\msun} limit in the Geneva tracks down to 0.15~{\msun} in the Padova tracks. Not only does this remove the previous restriction on modeling low-mass stars in young populations. More significantly, it allows Starburst99 to model stellar populations {\em of any age} up to the oldest cosmological ages with the best available stellar models.

\section{AGB models: first thermal pulse up to the planetary nebula}

We extended the published Padova tracks between 0.9 and 5~{\msun} to include thermally pulsing AGB stars. In this section we describe the adopted stellar properties in the AGB phase and their implementation in the Padova tracks.

The evolution of the AGB phase after the first thermal pulse has been the subject of numerous studies (e.g., Groenewegen \& de Jong 1993; Bl\"ocker 1995; Dorfi \& H\"ofner 1998; Ventura 2001; Wachter et al. 2002). Parameters of interest are the high mass-loss rate, variability in temperature and luminosity, and the time scale of this evolutionary phase. Observational guidance is provided, e.g., by Mira variables, where variability has been observed and attributed to thermal pulses (Wood \& Zarro 1981). The mass loss in this stage has been related to planetary nebula (PN) nuclei as well as to white dwarfs (WD) whose mass distribution peaks around 0.6 {\msun}. Therefore, WDs are considered the evolved descendants of AGB stars (Weidemann \& Koester 1983; Weidemann 1990). The progenitors of AGB stars are assumed to be $\ge 1.0$~{\msun} (Pottasch 1983; Jura 1990). The upper mass limit for AGB formation is somewhat uncertain but lies between 5 and 8~{\msun}, the latter value representing the transition to the formation of a core-collapse supernova. Renzini \& Voli (1981) noted that the mean mass-loss rate required to produce a typical PN was $\sim 3 \times 10^{-5}$~{\msun} yr$^{-1}$. They introduced the term {\em superwind} to describe such high mass-loss rates since they are much higher than those of RSGs described with the formula of Reimers (1975). Since then, many AGB stars (particularly infrared sources, OH/IR stars, and Mira variables) have been found to have mass-loss rates consistent with the superwind expectation (e.g., Knapp \& Morris 1985 in the Galaxy, and Wood et al. 1992 in the LMC).

Given the lower mass limit of the progenitor of $\sim 1.0$~{\msun}, it is clear that AGB stars are important for galactic ecology. The amount of gas mass converted into newly formed stars in this mass range is $\sim 25$\% of the total, depending on the initial mass function (IMF). Therefore, these stars become important for the total stellar mass loss in systems with a significant intermediate-age population. 

The central issue of the relation between the mass loss during the AGB evolution and other parameters, including the core mass of the star has received wide attention (e.g., Groenewegen \& de Jong 1993; Vassiliadis \& Wood 1993, Bl\"ocker 1995). The formulation of Groenwegen \& de Jong has been applied to the Padova tracks before (e.g., Girardi \& Bertelli 1998; Marigo 1998, Girardi et al. 2000). Since the Reimers (1975) mass-loss relation fails to reproduce AGB winds, an empirical approach relating $\dot M$ and other periodic properties in the thermally pulsing AGB phase was developed by Vassiliadis \& Wood (1993). We chose their models to develop a set of TIP-AGB parameters and add them to the evolutionary tracks from Padova.

Vassiliadis \& Wood (1993) published models with $Z = 0.004$, 0.008, and 0.02 for the mass range 0.9 - 5.0~{\msun}. Their models follow the evolution from the first thermal pulse in the EAGB phase up to the stripping of the core in the PN (or TIP-AGB) and WD phases. Their definition of the end of the AGB phase is the time when $T_{\rm{eff}}$ becomes higher by 0.3 dex than the value for a ``reference'' AGB star prior to blueward evolution. In our implementation of the Vassiliadis \& Wood models we terminated the evolution of the AGB phase in the final pulse before the last ejection of material at the lowest value of $T_{\rm{eff}}$. 

The models give $M_{\rm{c}}$, $T_{\rm{eff}}$, {\em L}, and $\dot M$ during the evolution from the first AGB pulse until the last pulse prior to the ejection of the nebula. We have interpolated these tracks in the range (0.9 - 5.0 {\msun}) to be consistent with the Padova tracks. The tracks were also re-sampled to cover the same evolution phases with a small number of points. The Padova tracks were re-sampled from the point where $Y_{\rm{c}} = 0$ up to the TIP-AGB phase (the pulse immediately prior to the lowest temperature) with 17 points over the tracks including the first thermal pulse already modeled in tracks. We used the Reimers formula to calculate the mass loss during the preceding RGB phase. 

A complication arises from the lack of AGB models for $Z = 0.0004$ and 0.05. It is currently unknown if and how the chemical composition affects the AGB evolution. In the absence of other evidence, we assumed the $Z = 0.004$ AGB models to be applicable to the $Z = 0.0004$ Padova tracks, and the $Z = 0.02$ models to the $Z = 0.05$ tracks. No further adjustment to the AGB models was made prior to the implementation. The main properties of the final set of tracks are summarized in Table~3. In this table we contrast the implementation of the Geneva and the Padova tracks with AGB stars in Starburst99. We list the range of abundances, mass range, the number of grid points, and the main evolutionary phases. Selected tracks with different chemical composition are shown in Fig.~2. In this figure we have highlighted the extension of the original Padova tracks towards cooler temperatures and higher luminosities with the addition of the TIP-AGB phase. As we will demonstrate below, this phase has significant effects on the predicted properties of stellar populations. In Fig.~3 we have plotted three tracks showing the AGB mass range at higher resolution for different chemical composition. Included in this figure are the Geneva tracks, the original Padova tracks, and the Padova tracks with the TIP-AGB phase. The comparison highlights the differences between the three sets of tracks for intermediate and low stellar masses. Lower metal abundances shifts all tracks to higher temperatures. In addition, the TIP-AGB correlates inversely with {\em L}: over most of the AGB mass range, more metal-poor stars are more luminous than their metal-rich cousins.

\section{Starburst99: Geneva vs. Padova tracks}

In this section we provide a comparison of synthetic populations generated with Starburst99 in order to identify similarities and differences introduced by the new set of tracks. A comparison with other synthesis codes will be performed in Section~7.

The model simulations in this and the following section were performed (unless otherwise stated) for a standard single stellar population of mass $10^6$~{\msun} with a Salpeter IMF between 0.1 and 100~{\msun}. We chose to use a low-mass cut-off of 0.1~{\msun} for the IMF, instead of the canonical value of 1~{\msun}. While this choice introduces an offset by a factor of $\sim 2.5$ with respect to the original Starburst99 models, it allows us to include low-mass stars below 1~{\msun} in the simulations, which would otherwise be excluded.

We used the fully line-blanketed atmospheres of Smith, Norris, \& Crowther (2002) for hot stars and ATLAS9$+$Phoenix models as adjusted by Lejeune, Cuisinier, \& Buser (1997; 1998) for stars of spectral type B and later.  The ATLAS9$+$Phoenix atmospheres were also used to assign spectral energy distributions (SED) to stars in the AGB phase. While this simplification is adequate for modeling SEDs, dedicated AGB spectra, such as those of Mouhcine \& Lan\c con (2002), are required for the synthesis of spectral lines.

The stellar atmospheres as well as most auxiliary tables in Starburst99 have chemical compositions tailored for the Geneva models. Since the Padova and Geneva tracks have different chemical composition at the low ($Z = 0.001$ and 0.0004 for Geneva and Padova, respectively) and the high ($Z = 0.04$ and 0.05, respectively) extremes, there is a slight mismatch in $Z$. We used the available data files (i.e., with $Z = 0.001$ and 0.04) for the most metal-poor and -rich Padova tracks. Consequently, the synthesized properties for these tracks have larger intrinsic errors than those for the $Z = 0.004$, 0.008, and 0.02 tracks where our abundances are consistent.

Previously published Starburst99 models were all performed with linear time steps. While this technique works fine for young ages, logarithmic time steps are more appropriate for populations with ages of order Gyr. Otherwise, either the time resolution at young ages would be unacceptably low, or the data volume would be unacceptably high. Consequently, the release of Starburst99 containing the Padova models offers the choice of linear or logarithmic time steps. In the following we will discuss representative model simulations with both the Geneva and Padova tracks for single populations. We also calculated corresponding models with continuous star formation but will not plot the results because of redundancy with the instantaneous models.

\subsection{Spectral energy distributions}

We have compared the stellar SEDs for the full age range. Throughout this paper we follow the convention to plot the Geneva, original Padova, and Padova$+$TIP-AGB models with solid, dashed, and dotted lines, respectively. Young populations with solar composition observed at 1, 3, 10, 50, and 100 Myr are shown in Fig.~4. SEDs produced with the Padova tracks are more luminous by up to 0.1 dex at most wavelengths. These differences are the result of higher intrinsic stellar luminosities in the Padova models for post-main-sequence evolutionary phases. 

There is a striking difference between the SEDs in the UV around 3 Myr. The excess emission in the Padova models is produced by hot W-R stars when coupled with extended model atmospheres. While this coupling is consistent when done for the Geneva models, it produces spurious results in the case of the Padova models. When W-R stars are important, the preferred approach is to use the Geneva tracks together with the extended atmospheres. If the Padova tracks are used, a minimalistic technique would be to couple them with the static ATLAS9 atmospheres. We will discuss the ionizing continuum in more detail in Section~6.4.

We selected five different ages of 0.1, 0.5, 1, 5, and 12 Gyr as representative for intermediate-age and old stellar populations. These models are plotted in Fig.~5. The contribution of the TIP-AGB stars to the red part of the spectrum is clearly visible: for models with ages $> 1$ Gyr (e.g., 5~Gyr), there is a strong difference between the Padova and Geneva tracks. The difference results from a combination of two effects. First, the Geneva tracks do not include low-mass stars below 0.8~{\msun}. Since our chosen IMF extends down to 0.1~{\msun}, the absence of these stars creates a luminosity deficit, which sets in at an age of a few Gyr, and becomes progressively stronger. Second, the implementation of the Geneva models in Starburst99 does not include the add-on tracks published by Charbonnel et al. (1996). Charbonnel et al. computed the horizontal branch evolution after the He-flash in low-mass stars. Horizontal branch stars around $\sim 1.5$~{\msun} contribute significantly to the near-IR luminosity beginning at an age of $\sim 1$ Gyr. They are the chief reason for the difference between Geneva and Padova at and after that epoch.

\subsection{Massive-star inventory}

The number of O stars derived from Padova models is the same as that from the Geneva models. This confirms the consistency between the two sets of tracks for the O phase in massive stars. The slight luminosity and temperature difference in the tracks is too minor to affect the stellar census in Starburst99.

Similarly, the supernova rates following from either set of tracks is essentially the same since the stellar lifetimes are very similar and the cut-off mass for supernova formation is the same.

In Fig. 6 we have plotted the W-R/O star ratios predicted by the Geneva and the Padova tracks. The ratios agree quite well, as expected from the similar W-R definitions in both sets. Recall that W-R stars are defined via their surface abundances and (for computational reasons) via their lower mass limit for formation. We apply the same definitions for both sets of tracks in Starburst99. Obviously, had we tied the W-R definition to $T_{\rm{eff}}$, the agreement in Fig. 6 would largely disappear because of the different temperature definitions. The small differences between the W-R/O ratios in Fig. 6 are caused by slightly different stellar lifetimes and numerical noise. As a reminder, W-R/O is strongly dependent on the chemical composition. A decrease in metals inhibits the formation of W-R stars.

W-R stars come in two subclasses: nitrogen-rich WN and carbon-rich WC stars. Their classification is tied to the surface abundances of He, C, N, and O. Details are in Leitherer et al. (1999). The WC/WN ratio is an important IMF tracer (Massey 2003). Fig. 7 suggests reasonable agreement between the predictions resulting from the Padova and the Geneva tracks. On average, the Geneva models lead to higher WC/WN ratios.

\subsection{Ionizing continuum}

We compared the ionizing fluxes from hot, massive OB stars resulting with the new sets of tracks. No significant differences are found for the hydrogen ionizing photons at any of the five chemical compositions. The fluxes shortward and close to the Lyman edge are predominantly contributed by OB stars, whose number densities are very similar for the two sets. W-R stars are negligible contributors for a standard population mix.

W-R stars do matter for the ionizing fluxes at shorter wavelengths in the He$^0$ and He$^+$ continuum below 504 and 228 {\AA}, respectively. The different $T_{\rm{eff}}$ definition in the Padova and Geneva tracks causes enormous differences in phases when W-R stars are present. This is illustrated in Figs.~8a and b for the number of photons capable of ionizing He$^0$ and He$^+$, respectively. In order to highlight how different model atmospheres interact with different tracks, we present four atmosphere/track combinations in the two figures. ``Atm~2'' refers to the plane-parallel Lejeune atmospheres in Starburst99, whereas ``Atm~5'' is the atmosphere package released by Smith et al. (2002), which assigns fully blanketed models accounting for stellar winds. 

Prior to the appearance of the first W-R stars (at $\sim 3$ Myr), the four atmosphere/track combinations predict He$^0$ ionizing photons which agree within a factor of 2 (Fig.~8a). During the W-R phase, the predictions diverge by up to a factor of 10. We consider the combination of the Geneva tracks and Smith et al. (2002) atmospheres as the currently best available models for W-R related parameters and use them as our baseline (solid line in Fig.~8a). When the Geneva tracks are coupled to the Smith atmospheres in Starburst99, we do not use the surface temperature but a hotter value from deeper in the atmosphere. This hotter temperature is the required input for the spherical extended atmospheres.  For comparison, linking the Lejeune atmospheres with the Geneva tracks leads to much lower ionizing fluxes from W-R stars because we use the lower surface temperatures in the evolution models and because the atmospheric fluxes of given temperature are slightly different from those of Smith et al. (long-dashed line). Contrasting these results with those from the Padova tracks, one recognizes that the higher $T_{\rm{eff}}$ in the Padova models {\em always} leads to higher ionizing fluxes, irrespective of the atmosphere. The small differences between the results for the static (dash-dotted) and expanding (short-dashed) atmospheres coupled with the Padova tracks are again caused by the different atmospheric fluxes.

 The interplay between the tracks and the atmospheres becomes even more dramatic at wavelengths below 228 {\AA} (Fig.~8b). O stars dominate the luminosity until about 3 Myr. The Smith et al. (2002) atmospheres use the WMBASIC wind models of Pauldrach et al. (2001) for O-star phases. These wind models predict 3 to 4 orders of magnitude higher luminosities than static ATLAS9 (Lejeune) models for both the Geneva and Padova tracks (solid and short-dashed lines in Fig.~8b). When W-R stars become important, Smith et al. rely on the CMFGEN atmospheres of Hillier \& Miller (1998). At solar and higher composition, the wind models predict recombination of He$^+$ in the W-R envelopes, and the emergent fluxes below 228 {\AA} are quite low (solid line). Since the wind recombination is not accounted for in the static ATLAS9 atmospheres, the prediction from the ``Geneva$+$Atm~2'' model (long-dashed) becomes unrealistically large. The behavior of the two Padova models is as expected: the dominant effect is the higher $T_{\rm{eff}}$, which boosts the He$^+$ luminosity by orders of magnitude. Because of the very high temperatures, recombination is less important, and the extended atmospheres predict only somewhat lower luminosities than the static models with the Padova tracks (short-dashed and dash-dotted lines).

The Padova models show a pronounced deficit of radiation around ages of 5~Myr when linked to the extended atmospheres. This depression results from the very short lifetime of a 25~{\msun} W-R star in the models. The 25 {\msun} track is unique in that it has a hot W-R star only during the last few thousand years prior to the termination of the track. In contrast, the 40 {\msun} model predicts the hot W-R phase to last for several hundred thousand years. Since the extended W-R atmospheres are only linked to the evolutionary models when both a minimum temperature and a minimum helium surface abundance are exceeded, there will be a period of a few $10^5$~yr around 5~Myr (corresponding to ZAMS masses around 30~{\msun}) without any stars that would satisfy the definition required by the atmospheres.

The graphs in Figs.~8a and b become very different at lower and higher chemical composition. This results from the combination of the dependence of the W-R formation on chemistry and the behavior of the W-R atmospheres on mass loss and metal abundance. Smith et al. (2002) give an extensive discussion of these effects as predicted for the Geneva models. As the Padova models are not primarily designed for modeling W-R stars, extending Fig.~8 to other chemical abundances would not add useful information to the discussion in the present paper.

\subsection{Panchromatic luminosities}

We compare the predicted luminosities at representative wavelengths in Fig.~9. As before, the Geneva, original Padova, and Padova$+$TIP-AGB models are presented. Fig.~9 shows the bolometric ($M_{\rm{Bol}}$), 1500 {\AA} ($L_{\rm{1500}}$), {\em B} band ($M_{\rm{B}}$), {\em V} band ($M_{\rm{V}}$), {\em J} band ($M_{\rm{J}}$), and {\em K} band ($M_{\rm{K}}$) luminosities. The passband definitions are the same as in Leitherer et al. (1999). All filters are in the Johnson (1966) system except for {\em R} and {\em I}, which are in the Cousins system (Bessell 1983; 1990). $L_{\rm{1500}}$ is the average of the luminosity between 1490 and 1510 {\AA}, and  $M_{\rm{Bol}}$ is referenced to a system in which the sun has $M_{\rm{Bol}} = 4.75$. (The latter zero point is equivalent to using the relation $\log(L/L_\sun) = -0.4 M_{\rm{Bol}} + 1.90$.) 

The overall trend in Fig.~9 suggests very good agreement between the three models until about 100 Myr when the first AGB stars form. The AGB phase lasts from 100~Myr to 2~Gyr and produces excess luminosity mostly in the IR. At epochs later than about 3~Gyr, the Geneva and Padova models diverge at all wavelengths, including $M_{\rm{Bol}}$. This is caused by the same effects already discussed in Section 6.1: the Geneva tracks do not account for the post-He-flash evolution in intermediate-mass stars and they lack low-mass stars altogether. As a result, the Geneva tracks lead to a light deficit when these stars become important.

The readers and users of Starburst99 should be aware of uncertainties inherent in $L_{\rm{1500}}$ for ages older than $\sim 3$~Gyr. After that epoch, UV-faint F and later-type stars provide the UV luminosity. The standard Kurucz model atmospheres are a poor match for the UV light of old populations (Lotz, Ferguson, \& Bohlin 2000). Efforts are underway to remedy the situation (Peterson, Dorman, \& Root 2001).

The results for different chemical abundances are qualitatively similar to those shown in Fig. 9 for solar composition. Changing the composition has its biggest effect in the RSG dominated phase when the evolutionary tracks predict strong variations of the RSG temperatures and lifetimes with abundance. We will test this prediction in Section~8.3.  

The fractional contribution of AGB stars to the luminosity in four passbands is reproduced in Fig.~10. Plotted is the magnitude increase by adding AGB stars relative to the prediction by the original Padova tracks. The four panels in Fig.~10 show $M_{\rm{Bol}}$, $M_{\rm{B}}$, $M_{\rm{V}}$, and $M_{\rm{K}}$ for three different compositions. As suggested by the trends in Fig.~9, AGB stars have minor influence on optical passbands, such as {\em B} and {\em V}, but can contribute up to 1 mag to near-IR filters. The contribution to the bolometric luminosity is 0.4 mag or less. This behavior is rather sensitive to variations of chemical composition as a result of the models of Vassiliadis and Wood (1993).

\subsection{Colors}

The color evolution is of particular observational interest, as it is a normalized quantity and does not depend on the (a priori unknown) star-formation history. We have plotted our results for ($B - V$), ($V - J$), and ($V - K$) in Fig.~11. In this figure, we include the color evolution for models with $Z = 0.02$, 0.008, and 0.004 in order to identify abundance effects. The overall trends are similar to those in Fig.~9. The Geneva and Padova models agree, except during periods when stellar phases dominate which are not well accounted for or are not included at all in one model set or the other. This is the case during the first few Myr when W-R stars contribute (not relevant to Fig.~11), during the AGB phase, and after $\sim 1$~Gyr when post-He-flash stars become important. The RSG dominated phase around 10 - 20~Myr differs in the Geneva and Padova tracks because of different lifetimes and temperatures. We will return to this issue in Section~8.

As a rule of thumb, main-sequence stars are always important contributors to optical and near-IR colors, except during the RSG dominated period between 10 and 20~Myr (the large depression in Fig.~11). AGB stars become comparable in their luminosity around 1~Gyr. At even older ages, RSGs provide the bulk of the luminosity. These arguments are of course dependent on the actual color and should be viewed as a rough guideline only. See also Charlot \& Bruzual (1991) and Maraston (1998) for a related discussion.   

Fig.~11 nicely illustrates how the colors are sensitive to chemical composition: more metal-poor stars are generally bluer than metal-rich stars (except during the AGB phase - see below). This is the joint result of bluer positions in the HRD and a steeper decrease of line-blanketing at shorter wavelengths for lower abundances. The RSG depression at $\sim 10$~Myr has a striking $Z$ dependence in both sets of tracks. There are fewer and warmer RSGs at lower chemical abundances, a theoretical prediction discussed by Cervi\~no \& Mas-Hesse (1994). We will perform a series of observational tests of this prediction in Section~8. 

The colors of AGB dominated phases are sensitive to chemical composition as well. The effect is most pronounced in the red and IR. This composition dependence is introduced by the Vassiliadis and Wood models, which predict a higher AGB luminosity at lower abundance. Consequently, AGB stars (which are very red) contribute more light at low {\em Z}. There is mild evidence for this effect from a comparison of Galactic and Magellanic Cloud AGB stars (Vassiliadis \& Wood 1993).

\subsection{Mass loss and mass-to-light ratio}

Starburst99 calculates quantities related to the mass and energy return in addition to the photon output. In the following we will use the mass-loss related properties as they follow from the table entries in the evolutionary tracks in order to be able to compare the tracks themselves. This differs from the recipe in Leitherer et al. (1999), where the stellar-wind properties were derived from a hybrid theoretical and empirical formula, independently of the stellar evolution models.

The mass loss by winds from all newly formed stars is plotted in Fig.~12.  The predictions following from either the Padova or the Geneva models are essentially identical until the appearance of the first AGB stars. The difference between the Geneva and Padova models around 100~Myr is caused by the lack of $\dot M$ entries for stars with $M < 12$~{\msun} in the Padova tables. The Padova$+$AGB tracks extend the prediction for $\dot M$ beyond $\sim 1$~Gyr, at which epoch Starburst99 would otherwise cease to be useful. AGB stars with their strong winds are the prime source for the mass return to the ISM for intermediate-mass stars. Nevertheless, the total mass lost in the AGB phase is minuscule when compared to the mass loss in earlier stellar phases. This can be appreciated in Fig.~13, where we have plotted the cumulative mass loss for the same stellar population. The AGB contribution is the incremental mass loss increase noticeable after about 3~Gyr. 

The corresponding mass-to-light ratios are in Fig.~14. We show the predictions for $L_{\rm{Bol}}$, $L_{\rm{V}}$, and $L_{\rm{K}}$. Other passbands behave accordingly. The three sets of tracks lead to very consistent results, except for those phases with missing stellar ingredients, as discussed before. We emphasize that the results in Fig.~14 do not take into account the removal of stellar material during the evolution of the population. In other words, {\em M} is constant and taken to be the initial total stellar mass. In practice, {\em M} decreases because some (or even most) of the mass lost by stellar winds can escape from the gravitational field of the stellar population and become unbound. Therefore, the mass-to-light ratios in Fig.~14 are lower limits to the astrophysical situation.

\section{Comparison of Starburst99 with other evolutionary synthesis codes}

How do the predictions of Starburst99 for the photometric properties of stellar populations with and without AGB stars compare to those from other synthesis codes? We surveyed the literature and collected the relevant data from those sources which either published the color evolution directly or allowed us the extract the photometry from the information provided in the publication. 

The literature sources used in this comparison are listed in Table 4, together with their principal ingredients. Included in this table are only those sources which specifically address the AGB evolution. Buzzoni (1989) released a widely used set of synthetic population properties but his models commence after 1~Gyr. Therefore we did not include them in our comparison. Col.~1 of this table gives the model designation. The evolutionary models are in col.~2. The source for the AGB evolution is in col.~3. In col.~4 we give the upper mass limit $M_{\rm{max}}$ included in the AGB evolution. The adopted SEDs and the photometric system are in cols.~5 and 6, respectively. In col.~7 we provide the literature source. Our point of reference is the Starburst99 implementation of the Padova$+$TIP-AGB models (this work). The second and third rows in Table~4 list the recent ``GALAXEV'' release by Bruzual \& Charlot (2003) who updated their earlier work by linking the new intermediate-resolution empirical library of Le Borgne et al. (2003). The GALAXEV models include the TIP-AGB evolution for both the original and the Girardi versions of the Padova tracks. Except for the use of the empirical library, the ingredients in the GALAXEV models are closest to those of Starburst99 among the entries in Table~4. ``Spectral'' (V\'azquez, Carigi, \& Gonz\'alez 2003) incorporates both the Padova {\em isochrones} as released in the paper of Girardi et al. (2000) or the Padova (1994) {\em isochrones} (Bertelli et al. 1994), respectively\footnote{The Padova isochrones should not be confused with the tracks. The isochrones contain $T_{\rm {eff}}$ and $L$, and these quantities are directly used by Spectral for the synthesis}. The Spectral ingredients include a very approximate model for TIP-AGB stars from Boothroyd \& Sackmann (1988) and Groenewegen \& de Jong (1993) in the 1994 and 2000 isochrones, respectively. Mouhcine \& Lan\c con (2002) published dedicated AGB evolution models, based on the earlier work of Wagenhuber \& Groenewegen (1998). Maraston's (1998) models are based on the fuel-consumption theorem (Buzzoni 1989) and incorporate the stellar evolution models of Castellani, Cheffi, \& Straniero (1992). These evolution models are for the assumption of no overshooting, making them a useful test for this particular ingredient. For the AGB phase the models from Bl\"ocker and Sch\"onberner were used. Finally, we include the ``PEGASE'' code of Fioc \& Rocca-Volmerange (1997) which is complete for the full AGB phase, including the post-AGB evolution as calculated by Sch\"onberner (1993) and Bl\"ocker (1995).  

In addition to differences in the adopted evolution models, the eight codes in Table~4 consider quite different mass ranges for the AGB phase. We note that the lower boundary mass is only of academic interest as long as it is below 1~{\msun} because less massive stars have not had time yet to reach the AGB phase since the first stars formed in the universe. The upper boundary mass is more relevant. Maraston (1998) assumes the highest value, i.e., 8~{\msun}, whereas most other work uses 5~{\msun}. The Starburst99 value is 5~{\msun}. Fortunately, these differences have little effect on the colors of the population. Massive AGB stars, while luminous, are quite rare because of the IMF weighting, and they have shorter life times. Therefore the relevant masses are between 3 and 5~{\msun}, which are covered by all entries in Table 4.

The synthesis codes in Table~4 fall into two groups with respect to their assumptions for the SEDs. Fioc \& Rocca-Volmerange (1997), Mouhcine \& Lan\c con (2002), and Bruzual \& Charlot (2003) incorporate empirical library stars to reproduce AGB stars, the other models (including Starburst99) rely on stellar atmospheres. The SED entries in Table~4 are high-level guidelines only. Often the authors use an elaborate mix of several libraries with cross-calibrations. The salient point is that one group of entries in Table~4 is primarily based on empirical, and the other on theoretical SEDs. Finally, the photometric systems can differ. Some predictions are in the original Johnson (1966) system, whereas others are in the Johnson-Cousins system as defined by Bessell (1983; 1990) and Bessell \& Brett (1988). While the different filter profiles and zero points are small in most filters, they do matter for the {\em R} and {\em I} passbands.

None of the listed codes uses exactly the same ingredients as Starburst99. Nevertheless, by comparing the different model predictions on a relative basis, we can identify the trends introduced by different assumptions and attempt to verify the consistency of Starburst99 with other work.

In Fig.~15 we have plotted the color evolution in the ($U - B$), ($B - V$), ($V - R$), and ($V - K$) colors for solar abundance. For the sake of clarity we plotted Starburst99 and GALAXEV in the left panels, and the remaining models in the right panels. (Note that Mouhcine \& Lan\c con 2002 did not publish optical colors. Their IR results are discussed separately further below.) Except for random numerical noise, Starburst99 and GALAXEV have no systematic differences by more than $\sim 0.05$~mag in the optical colors, and $\sim 0.2$~mag in the near-IR. We do not detect significant differences with respect to either the Padova (1994) or the Girardi (2000) implementations in GALAXEV. Although this agreement was hoped for because of the similar inputs, it is nevertheless a gratifying result.

The comparison with the predictions of Spectral, Maraston (1998), PEGASE, and Mouhcine \& Lan\c con (2002) is in the right panels of Fig.~15. We note the generally very good agreement in ($U - B$) and ($B - V$) at any age and in ($V - R$) and ($V - K$) outside the AGB phase. When AGB stars dominate, the predictions differ by up to one magnitude. The differences can be understood from the earlier discussion of the different stellar tracks and AGB models used by the different codes. The predictions of Maraston (1998) in particular differ because of their use of a fully independent set of stellar evolution models. The agreement between the predictions of the other codes does not necessarily suggest those models are preferable; the codes responsible for those predictions are simply using the same evolutionary tracks. Furthermore, Maraston's ($V - R$) colors are in the original Johnson (1966) system. The {\em R} filter has a pivot wavelength of 0.69~$\mu$m, whereas the Cousins {\em R} is at 0.64~$\mu$m. This leads to redder ($V - R$) colors in the Maraston models. The Padova isochrones in their implementation in Spectral lead to very high AGB luminosities. This is the result of the adopted temperature-luminosity relation hard-coded in the Padova isochrones release.

Next, we turn to the near-IR colors. Since some literature sources provide only the near-IR colors, others only the optical colors, and some a mix of both wavelength ranges, we are presenting the near-IR results separately in Fig.~16. This figure serves a dual purpose. First, we are comparing the AGB sensitive near-IR colors for different codes, and second, we are evaluating the influence of the heavy-element abundance. The comparison at solar chemical composition is consistent with the previous results at shorter wavelengths. Starburst99 and GALAXEV are in excellent agreement. The comparison with the models of V\'azquez et al. (2003), Fioc \& Rocca-Volmerange (1997), and Mouhcine \& Lan\c con (2002) is somewhat less favorable but nevertheless still acceptable. The trend with composition in the AGB phase is opposite to that seen at older ages: lower abundances lead to redder colors. 

Finally, we provide a comparison of the mass-to-light ratio predicted by Starburst99 using the Padova$+$TIP-AGB tracks and by those sources in Table~4 which have published this quantity. The mass-to-light ratios in Fig.~17 were obtained under two assumptions for the mass evolution in Starburst99. In one case we adopted a constant mass equal to the initial total stellar mass, and in the other we allowed for a continuous mass decrease according to the mass loss prescription in the stellar-evolution models. We are distinguishing between these two cases for a fair comparison with other models, some of which were computed only with (Maraston 1998) and some only without (Fioc \& Rocca-Volmerange 1997) mass loss.  The GALAXEV models consider both cases. Fig. 17 suggests excellent agreement overall. Starburst99 and GALAXEV are the best matches for either mass-loss model. As expected, Starburst99 is in better agreement with Fioc \& Rocca-Volmerange when M is constant in Starburst99 (upper panels). On the other hand, the models of Maraston agree better with Starburst99 for non-conservative evolution (lower panels). Again, we note the impact of the different stellar evolution models used by Maraston and Starburst99.

\section{Testing the predictions of Starburst99 for intermediate-age and old populations}

In this section we will perform basic performance tests of our models by comparing the predicted colors to sets of relevant observational data. As before, we will ignore very young, ionizing populations since the new set of evolutionary models is not optimized for these populations, and extensive tests of Starburst99 applied to hot stellar populations have been reported in the literature. Three age regimes will be explored: $\sim 10^{10}$~yr, as traced by old globular clusters, $\sim 10^{8}$~yr observed in super star clusters, and $\sim 10^{7}$~yr, the age regime of RSGs.

\subsection{Globular clusters in NGC~5128}
 
Peng, Ford, \& Freeman (2004a, b) performed an in-depth study of the globular cluster system surrounding NGC~5128. It is not our goal to repeat their excellent analysis. Rather, we will take advantage of their photometric database of 215 confirmed globular clusters with radial velocities and compare the measured ($U -  B$), ($B - V$), ($V - R$), and ($V - I$) colors to those predicted by Starburst99. This comparison is done in Fig.~18 for ($B - V$) and ($V - I$). We constructed two-color diagrams for the other colors as well, with consistent results. These diagrams are not reproduced here.

The theoretical colors in Fig.~18 cover the age range from 0 to 10~Gyr, with tick marks highlighting 0.1, 1, 5, and 10~Gyr. Models for three chemical abundances are plotted: $Z = 0.0004$, 0.004, and 0.02. We restrict our discussion to models using the Padova$+$TIP-AGB tracks since the Geneva tracks are known to fail at those old ages. The models illustrate a textbook example of the notorious age-metallicity degeneracy (Worthey 1994): the metallicity and the age vectors are almost perfectly parallel at old ages, making it impossible to separate the two parameters in the absence of additional information. The data points are the entries from the catalog of Peng et al. (2004a). They are not corrected for foreground reddening. Peng et al. (2004b) favor a reddening of $E(B - V) = 0.115$, and the corresponding shift of the data points is indicated by the reddening vector. Note that the vector is parallel to the direction of the metallicity and age evolution.

The colors predicted by Starburst99 coincide with the ($B - V$)-($V - I$) plane defined by the observations. As discussed by Peng et al. (2004b), the globular cluster system of NGC~5128 consists of two well-defined age and metallicity groups, one having ($B - V$)$_0$ = 0.71 and ($V - I$)$_0$ = 0.89, and the other 0.88 and 1.10, respectively. (The subscripts indicate dereddened colors.)  The mean colors of the two cluster groups are in excellent agreement with, e.g., the age $\ge 10$~Gyr for $Z = 0.004$ and the age $\le 10$~Gyr for $Z = 0.02$ models. Of course other age/metallicity combinations in the models reproduce the data as well. The main point is that the models are an excellent match to the data, given the restrictions imposed by the age-metallicity degeneracy.

\subsection{Super star clusters in NGC 4038/39}

NGC~4038/39 ("The Antennae") are the archetypal interacting galaxies whose newly formed system of super star cluster is interpreted as an important piece of support in favor of the merger-formation hypothesis for elliptical galaxies (Schweizer 1998). Whitmore et al. (1999) obtained deep HST WFPC2 photometry of NGC~4038/39 and identified distinct age groups ranging from very young ionizing clusters, over intermediate-age ($\sim 10^{8}$~yr), to old globular clusters. We can utilize their published photometry for a model comparison with an emphasis on intermediate ages. We proceed as in the case of NGC~5128, i.e., we constructed two-color diagrams for all published passbands but we will reproduce only the ($V - I$) vs. ($B - V$) diagram (Fig. 19). The lay-out of Fig. 19 is identical to that of Fig.~18. We show the same sets of models, and we include the published data points without the appropriate reddening correction. The latter is rather substantial for young clusters because of the significant internal reddening of  $E(B - V) = 0.3$ in NGC~4038/39 (Whitmore et al. 1999).  

Whitmore et al. (1999) detected 11 very red clusters which they interpreted as old globular clusters from the original pre-merger galaxies. Comparison with the models in Fig. 19 confirms their suggestion. The 11 clusters occupy the same parameter space as the globular clusters of NGC 5128 in Fig. 18. Any of the three plotted age sequences would be able to reproduce their colors with a suitably chosen age/metallicity combination.

The intermediate-age clusters fall into the age range 0.25 to 1 Gyr according to Whitmore et al. (1999). 25 such clusters were identified by Whitmore et al. and have been added to Fig.~19. Almost all super star clusters in this age group are clustered around ($B - V$) $\sim 0.25$ and ($V - I$) $\sim 0.4$. This is the age when the first AGB stars influence the colors in our models. The models are not affected by an age-metallicity degeneracy in this age range ($\sim 200$~Myr) because of the dependence of the AGB phase on chemical composition. It is important to understand the physical reasons for the vertical loops at ($B - V$) $\sim 0.0$ to 0.3 in Fig.~19. The loop for the $Z_\odot$ models is caused by RSGs, which are much less important lower abundances. The redder ($V - I$) at higher {\em Z} reflects the lower $T_{\rm{eff}}$ of RSGs for higher abundances. In contrast, the loops in the $Z = 0.004$ and 0.0004 models are caused by AGB stars, which are predicted to have higher {\em L} at lower {\em Z}. Since both RSGs and AGB stars are very red, they are much more important for red than for blue stars. Therefore, ($B - V$) does not show these {\em Z} dependences, and the distinct loops form in Fig.~19. NGC~4038/39 has solar chemical composition, and the corresponding model sequence is a rather good match to the data. Conversely, the models with sub-solar chemical composition are clearly excluded. Of course, this prediction calls for independent verification with more data, given the results with find for RSGs in Section 8.3. Nevertheless, we can state that models with appropriate abundance do in fact reproduce the observed cluster colors.

The youngest clusters in the sample of Whitmore et al. (1999) are characterized by the bluest colors and their association with HII regions. These 50 objects in Fig.~19 exhibit the largest dispersion. The most likely cause is strong contamination by nebular H$_\beta$ and [O {\small III}]. The {\em V} photometry was obtained by transforming the original WFPC2 F555W data into the Johnson system. The F555W filter encompasses several strong nebular lines, including the mentioned H$_\beta$ and [O {\small III}]. In contrast, the F439W (used in lieu of {\em B}) and the F814W (used in lieu of {\em I}) filters do not suffer from such contamination. Therefore the data points representing the youngest clusters in Fig.~19 must be corrected towards bluer ($B - V$) and redder ($V - I$), i.e., towards the upper left. In addition, the clusters are affected by strong internal reddening, which is not accounted for. A quantitative interpretation of the youngest clusters is not our goal. Instead we will test the model predictions for stellar populations with ages of tens of Myr with blue compact dwarf galaxies observed in the IR.

\subsection{The red stellar population of blue compact dwarf galaxies}

Our next test will specifically address the validity of our evolutionary synthesis models for younger age ranges when RSGs are important. This is roughly the epoch between 8 and 30~Myr (see Fig.~11). A stellar population observed in this phase can have near-IR colors that are comparable to those of very old populations. Doubts about the validity of the stellar evolution models used in the synthesis were raised by Origlia et al. (1999), who were unable to reproduce the colors of RSG dominated populous clusters in the Large Magellanic Cloud with Starburst99. It is therefore of interest to evaluate if the use of the Padova tracks leads to a significant improvement.

Since the LMC cluster sample used by Origlia et al. (1999) suffers from small-number statistics, we decided to use published near-IR photometry of blue compact dwarf (BCD) galaxies as test beds. BCDs are the archetypal starbursts: they were among the galaxies first suggested to experience a ``burst (or flash) of star formation'' by Searle, Sargent, \& Bagnuolo (1973). BCDs are considered to be strongly starbursting galaxies powered by ionizing stars and an underlying population of red stars whose age is still under discussion (e.g., Aloisi, Tosi, \& Greggio 1999; Thuan, Izotov, \& Foltz 1999). There is, however, convincing evidence for a significant number of RSGs in those objects with well-established color-magnitude diagrams (e.g., Schulte-Ladbeck et al. 2001).

A literature search for existing near-IR photometry of BCD galaxies resulted in the selection of the studies Thuan (1983), Doublier, Caulet, \& Comte (2001), Noeske et al. (2003), and Telles (2004). These authors obtained {\em J}, {\em H}, and {\em K} large-aperture photometry of a statistically significant number of BCDs. Thuan's work was done with a classical InSb photometer, but through a wide aperture of at least 8'', which is large enough to encompass the whole galaxy. The other three studies used modern two-dimensional detectors, ensuring coverage of the entire underlying galaxies.  The photometry in all four studies is both sensitive to the young starburst, the surrounding field consisting of potential earlier starburst episodes, and the older underlying population. Therefore any RSGs present in these galaxies will affect or even dominate the photometry.

The photometry values collected from the four papers are in Table~5. This table gives the identifier in col.~1, where KN, VD, TT, and ET stand for the initials of the lead authors. Col.~2 lists the galaxy names. The near-IR colors ($J - H$) and ($H - K$) as well as their corresponding quoted errors are in cols.~3 - 6. The table has a total of 79 galaxy entries, although four are non-unique. The photometry was taken as given in the publications; no correction for foreground reddening was applied by us. This correction was performed by Thuan (1983) but not in the other three studies. However, the corrections would be less than 0.03~mag in all cases, which is totally negligible in view of other uncertainties. 

The data points from Table~5 are plotted in a color-color diagram reproduced in Fig.~20. The error crosses are the average of the individual photometric errors of each study, off-set in order to minimize confusion. Before comparing the data to synthetic colors, two corrections need to be addressed. First, the {\em J}, {\em H} and {\em K} bands encompass several strong nebular emission lines, such as Pa$\beta$, which can be strong in the presence of ionizing stars (Dale et al. 2004). Vanzi, Hunt, \& Thuan (2002) obtained near-IR spectra of several galaxies for which they had photometry and determined the color corrections that need to be applied to the {\em J}, {\em H}, and {\em K} data. Their average corrections are $+0.09$ and $-0.05$~mag in the ($J - H$) and ($H - K$) colors, respectively. These values include both the line and the continuous emission. The continuous nebular emission (free-free and bound-free) in the {\em J}, {\em H}, and {\em K} bands is much more significant, but it is accounted for in the synthetic colors and does not have to be removed for a comparison. Therefore, the pure line emission correction to be applied to our sample should be much smaller than the above values. We conclude that this effect is negligible. 

We also plotted the reddening vector for $A_{\rm{V}} = 0.25$~mag in Fig.~20. This corresponds to the average {\em internal} reddening of BCDs as derived from the optical Balmer decrement (Thuan 1983). The dispersion, however, can be large. For instance, Tol~0610-387 has essentially zero internal reddening, whereas Mrk~996 has a large value of $\sim 0.2$ in ($H - K$). The overall effect of both line emission and internal reddening is that the bulk of the galaxies could move to bluer (smaller) values of  ($J - H$) and ($H - K$) by less than 0.5~mag, with a few individual objects moving close to the (0,0) point in Fig.~20.

The synthetic colors predicted for a single stellar population by Starburst99 and the Geneva tracks are plotted in Fig.~20 as the short-dashed, long-dashed, and solid lines for $Z = 0.02$, 0.004, and 0.0004, respectively. The full symbols are tick marks to indicate ages of 5, 10, 25, and 100~Myr. Before proceeding with a discussion of the implications, we introduce Figs.~21, 22, and 23. These figures are identical to Fig.~20 in their data contents but show different model predictions. In Fig.~21, we have plotted the predictions from the Padova+TIP-AGB tracks with $Z = 0.02$, 0.004, and 0.0004. Figs.~22 and 23 are the counterparts of Figs.~20 and 21, respectively, in that we have substituted the predictions for a single starburst by those for continuous star formation.

Figs.~20 and 21 suggest reasonable agreement between the bulk of the data and the colors for solar chemical composition. The Geneva models appear to do marginally better, but this is mainly caused by the 10~Myr model point. The predicted lifetime of this red loop is quite short so that the 25~Myr model point should be a better indication of the location of the {\em majority} of the data. In that respect, the Padova and Geneva models are quite similar. Of course, a few of the observed BCD will in fact have ages of about 10~Myr with little internal dust. These objects would require the extreme Geneva data point. We terminated the synthetic models at an age of 100~Myr. Higher ages are unrealistic, as BCDs are defined via their emission lines and blue colors. 100~Myr old starbursts would not be classified as BCDs anymore.  

The detailed star-formation history of BCDs is a subject of debate. Dwarf galaxies are known to have had rather complex star-formation histories, with periods of quiescence and intermittent bursts of star formation (e.g., Greggio et al. 1998). Depending on the burst frequency, the effective star formation may be closer to the continuous than to the instantaneous case. This is addressed in Figs.~22 and 23. As expected, the imprint of the RSGs is more diluted, and the colors are more degenerate than for a single stellar population. If the star-formation histories in BCDs were constant, the comparison between the data and the models would force us to postulate ages far in excess of 100 Myr. While this would not necessarily be in conflict with observational selections (ionizing stars are continuously replenished), the associated gas consumption would become entirely unreasonable. BCDs do not constantly form stars over $\sim 1$~Gyr. Therefore we conclude that the appropriate star-formation scenario is between those plotted in Figs.~20/21 and 22/23, but most likely much closer to the instantaneous case of Figs.~20 and 21.

Does this suggest consistency between the observed and synthetic colors? The only track in Fig.~20 or 21 that matches the data points is the one at solar chemical composition. The other tracks at lower abundance are significantly bluer and fail to reproduce the observed colors regardless of the assumed age and reddening correction. Only the solar models produce RSGs in large enough numbers and with sufficiently low $T_{\rm{eff}}$ to intercept with the band delineated by the reddening corrected data points. A literature survey suggests that the approximate average oxygen abundance of our sample is 20\% solar. Therefore solar composition is entirely excluded, and the applicable models are those having Z = 0.004. Once we are forced to compare the data to the Z = 0.004 tracks, we arrive at the same conclusion as Origlia et al. (1999): {\em The predicted colors of evolutionary models for metal-poor populations with a significant RSG component are incorrect}. This conclusion is unchanged for either the Padova or Geneva tracks. 

Our results echo those of Massey \& Olsen (2003), who demonstrated that both the Geneva and Padova evolution models fail to predict the location of RSGs and the Large and Small Magellanic Clouds. First, the models are too warm at any temperature, and second they do not reproduce the smooth decrease in effective temperature with increasing luminosity. The higher luminosity RSGs are invariably of cooler temperatures, which is just the opposite of the predicted behavior. The failure of stellar evolution models, in particular at sub-solar chemical composition, implies that any spectral synthesis model for metal-poor stellar populations with RSGs is doomed to fail as well. The evidence of failure at solar chemical composition is much weaker, if present at all. 

The disagreement does not entirely come as a surprise. Standard stellar models are still based on the half-century old mixing-length approximation for describing the convective energy transport. One assumes a fixed value of the ratio of mixing length to pressure scale height that directs convection in the same manner in all different environments. The mixing-length parameter primarily governs the stellar radius and is derived from a solar calibration. Even if it can be justified in Sun-like stars, it may break down in environments that are different from that of the Sun. Such environments may include not only chemically different stars but also different evolutionary stages of the same star. Although there are efforts to find more realistic descriptions of convection (e.g., Yi 2003), there is no comprehensive set of stellar models based on a more mature convection prescription. 

Another source of uncertainty is the amount of convective overshooting. Overshoot is the penetrative motion of convective cells, reaching beyond the convective core as defined by the classic Schwarzschild criterion. Its effects on stellar evolution are ubiqitous, most notably on the ratio of lifetimes spent in the core hydrogen-burning stage and in the shell hydrogen-burning stage. Yi et al. (2000) performed an analysis of such effects in isochrone and spectrum synthesis. By definition, overshooting is effective only for those stars whose high masses lead to the formation of a convective core. Therefore it affects only young populations, producing bluer spectra (Yi 2003).

Finally, evolution models with rotation are just beginning to become available (Maeder \& Meynet 2003). The new models demonstrate the significance of rotation: The additional helium brought near the H-burning shell by rotational mixing and the larger He-core both lead to a less efficient H-burning shell and a smaller associated convective zone. Therefore, the stellar radius of rotating stars will inflate during the He-burning phase. These models account for the formation of more RSGs at low $Z$, with a blue-to-red supergiant ratio in better agreement with observations. It may very well be that rotation is a dominant effect and controls the interpretation of the spectra of high redshift galaxies and the early nucleosynthesis.

Frustratingly, mixing length, overshooting and rotation are the likely culprits for the failure of the evolution models, but they cannot be predicted from first principles and remain adjustable parameters. Only empirical calibrations can provide constraints.

Origlia et al. (1999) attempted to derive such an empirical calibration. They developed evolutionary synthesis models from Starburst99 and applied them to near-IR spectral features observed in the spectra of young Magellanic Cloud clusters. The temporal evolution of the first and second overtones of CO at 2.29~$\mu$m (2-0 band head) and 1.62~$\mu$m (6-3 band head), and of the $(U - B)$, $(B - V)$, and $(J - K)$ colors were investigated. Origlia et al. utilized star clusters whose ages were known independently from spectroscopy, allowing them to break the age-related degeneracy. They found that the evolutionary models of RSGs with sub-solar chemical composition are not reliable for any synthesis of the temporal evolution of IR features. While this mirrors the results of Massey \& Olsen (2003), the integrated light analysis of single clusters allowed them to place constraints on the temperature and the fraction of time spent in the red part of the (HRD) by massive stars during their core-helium burning phase. The study of Origlia et al also addressed the question if a different choice of the IMF could improve the agreement, with a negative answer. The near-IR spectral features from RSGs are practically independent of the IMF. The negligible dependence on the IMF simply follows from the fact that RSGs provide both the line flux and the adjacent continuum. Therefore, the IR indices sample a relative small stellar mass interval and become IMF independent.

The key result of Origlia et al. (1999) is that the evolution models fail not only in their prediction of the temperature but also of the lifetimes spent in the cool RSG phase. Origlia et al. derived a set of empirically calibrated spectrophotometric models by adjusting the RSG parameters so that the properties of the observed templates were reproduced. Specifically they decreased the effective temperatures and increased the lifetimes of the blue core-helium burning phases. Their study suffers from two major limitations: (i) The metallicity range is restricted to that of the Magellanic Clouds. Yet, we expect the most dramatic shortcomings of stellar evolution models at even lower metallicity. (ii) An integrated light study is unable to break the mass degeneracy. Origlia et al. had to assume that the same correction factors apply to RSGs of all masses, which is obviously a very drastic assumption with little theoretical or empirical justification.

Therefore we are forced to advice against using Starburst99 to model stellar populations at sub-solar chemical abundances whose light contains a significant RSG component.

\section{Conclusion}

We have enhanced Starburst99 by implementing a new set of evolutionary tracks which extends the useful modeling space of this synthesis code to ages older than $\sim 1$~Gyr and masses below 0.8~{\msun}. The prior restrictions in Starburst99 were imposed by using the Geneva models, whose are optimized for young massive stars. In contrast, the newly added Padova tracks are more complete and have more detailed input physics at intermediate and low masses. We augmented the Padova tracks with dedicated models for AGB stars in order to complete the evolution of intermediate-mass stars until the final thermal pulse prior to entering the PN phase. With these enhancements, Starburst99 allows the calculation of synthetic population properties for virtually any astrophysical situation of interest.

We discussed in this paper the preferred domains where either the Geneva or the Padova models are a better match to observations. The former are preferred for young, ionizing populations whereas the latter are the models of choice when intermediate-age and old population dominate. The implementation in Starburst99 offers users the choice of performing synthesis calculations with both sets of tracks in order to help make an educated decision as to which are the preferred tracks.

Comparisons with published synthetic properties by other groups suggest consistency once differences in the adopted input physics are taken into account. In this respect, Starburst99 resembles GALAXEV most closely, and it agrees very well with the predictions of that code. 

We performed a basic sanity check of our models by comparing the predicted colors to observations of old, intermediate-age, and young clusters. The agreement is excellent for ages above $10^8$~years, which includes the full AGB phase. When RSGs dominate, however, models and observations at subsolar chemical composition disagree. This discrepancy is caused by incorrect parameters of RSGs in the stellar evolution models.

All model calculations presented in this paper can be repeated by using the on-line version of Starburst99 at http://www.stsci.edu/science/starburst99/. The upgrades described here are released as version 5.0. In addition, v5.0 has the following changes with respect to the currently available v4.0.

-- Both linear and logarithmic time steps can be selected. The latter are a better compromise between sufficient time resolution and data volume management.

-- We replaced the previously implemented Johnson R and I filters with the more widely used Cousins R and I filters. Since the Cousins filters have shorter effective wavelengths, colors in these two passbands tend to be bluer when calculated with Starburst99 v5.0.

-- The on-line version of Starburst99 now offers more choices for the assumed IMF. While the previous version supported a single power law, we implemented multiple power laws in v5.0. Therefore, the commonly used parameterization of Kroupa (2002) can be adopted as input.

We encourage the community to take advantage of the improvements whose come with the new release of Starburst99.



\acknowledgments

We are grateful to Miguel Cervi\~no, Cesare Chiosi, Daniel Schaerer, and Eva Villaver for providing useful comments during various phases of this project. Edu Telles kindly made his unpublished photometric data available to us. This work was supported by HST grant GO-9116 from the Space Telescope Science Institute, which is operated by the Association of Universities for Research in Astronomy, Inc., under NASA contract NAS5-26555.



\appendix

\clearpage

\begin{figure}
\plotone{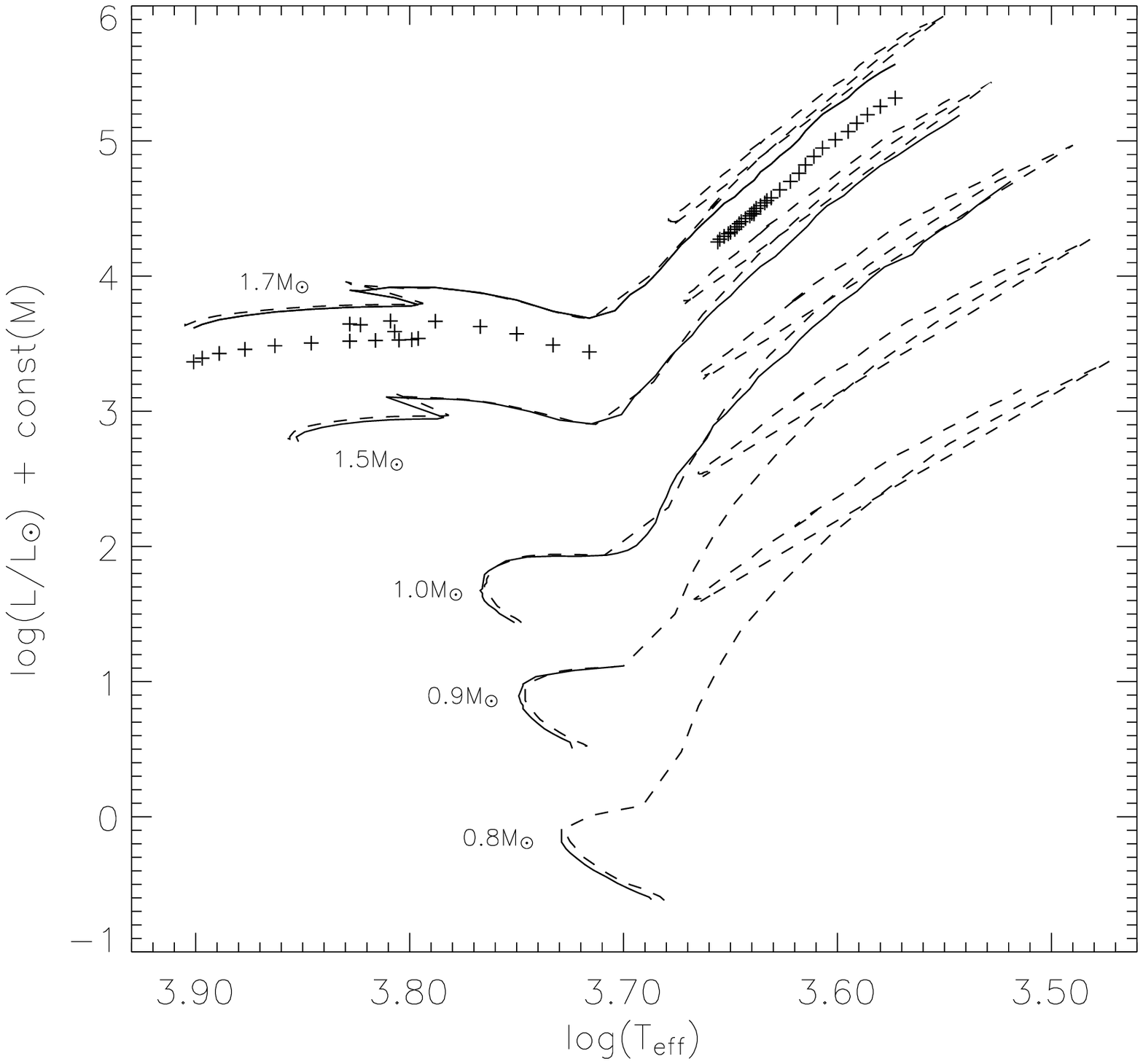}
\caption{Geneva (solid) and Padova (dashed) tracks for solar composition used in Starburst99 showing the transition from the intermediate to the low-mass regime. The tracks are shifted in $\log L$ from their original value by 0, 0.9, 1.6, 2.1 and 2.7 dex for $M = 0.8$, 0.9, 1, 1.5, and 1.7~{\msun}, respectively. An additional Geneva track for 1.7~{\msun} is plotted to show the transition from stars with and without degenerate helium core (crosses, shifted by 2.45 from its original value) in the Geneva tracks. 
\label{fig1}}
\end{figure}


\begin{figure}
\plotone{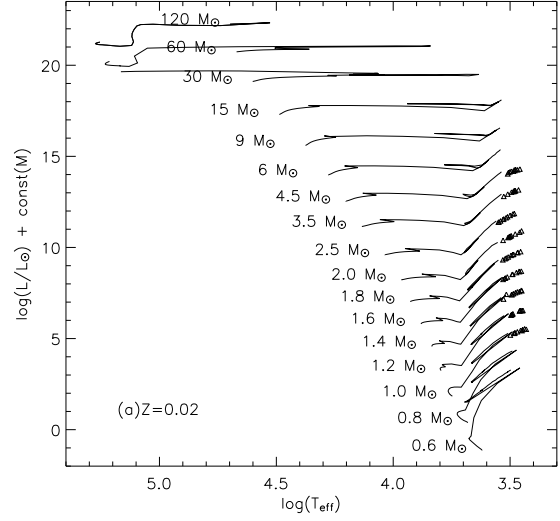}
\caption{The Padova tracks with the TIP-AGB for (a) $Z = 0.02$, (b) $Z = 0.008$, and (c) $Z=0.004$. The tracks are shifted in $\log(L/L_\sun)$ by 0, 1, 2, ... 16 dex from their original value of $M_{\rm{i}} = 0.6$, 0.8, 1.0 . . . 120~{\msun}, respectively, to improve clarity. AGB phases are identified by open triangles.\label{fig2}}
\end{figure}


\begin{figure}
\plotone{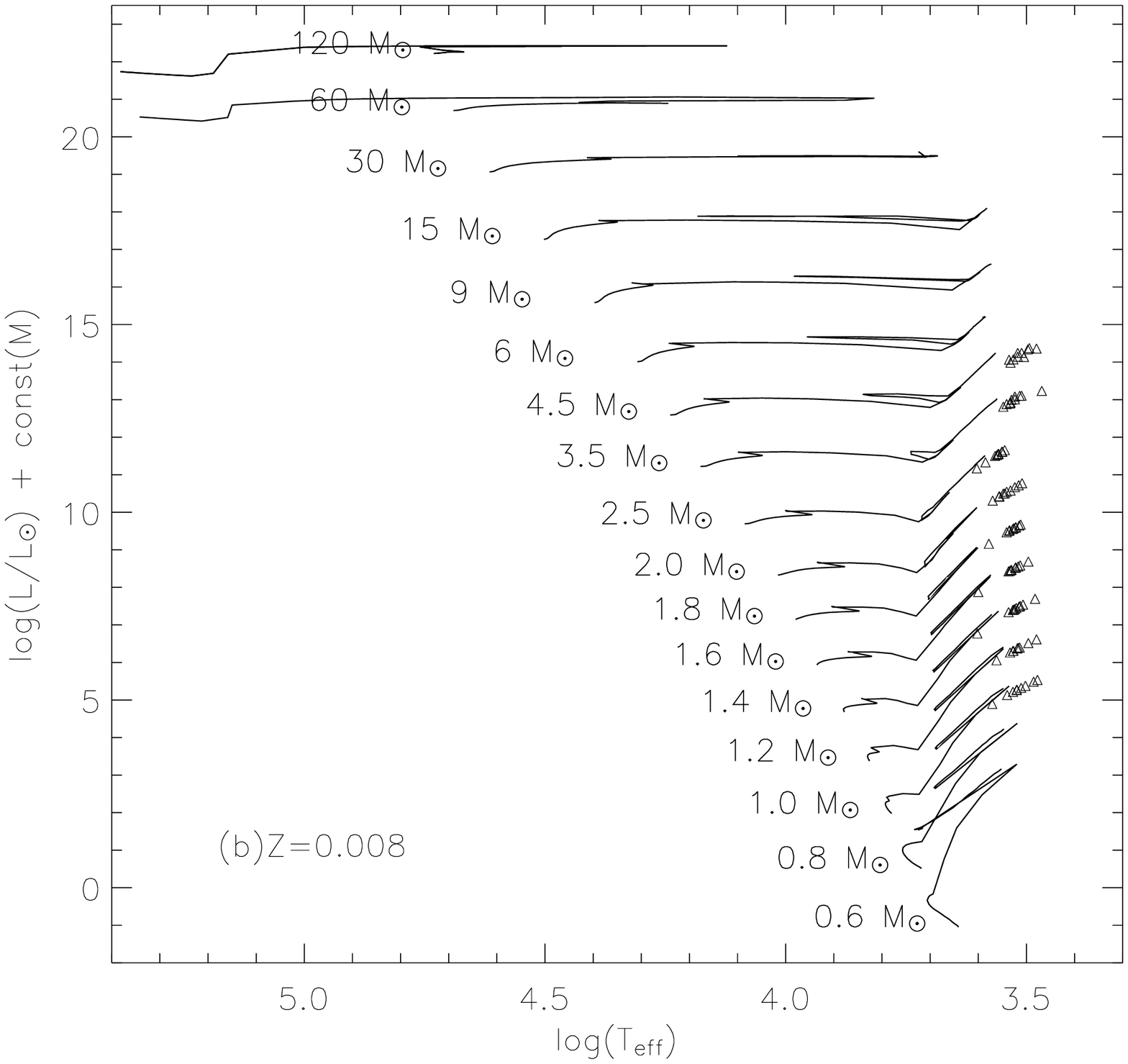}
\end{figure}


\begin{figure}
\plotone{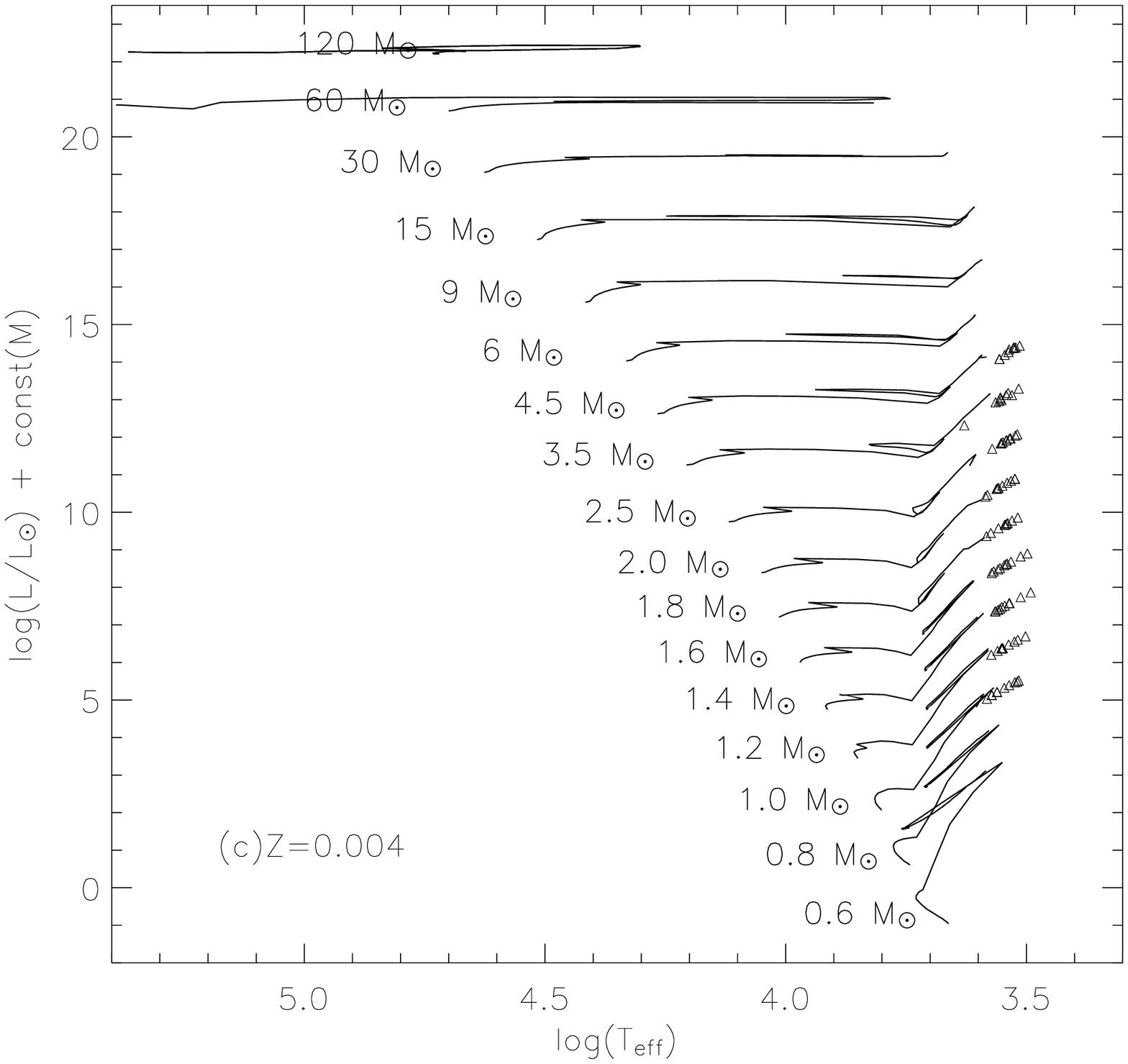}
\end{figure}


\begin{figure}
\plotone{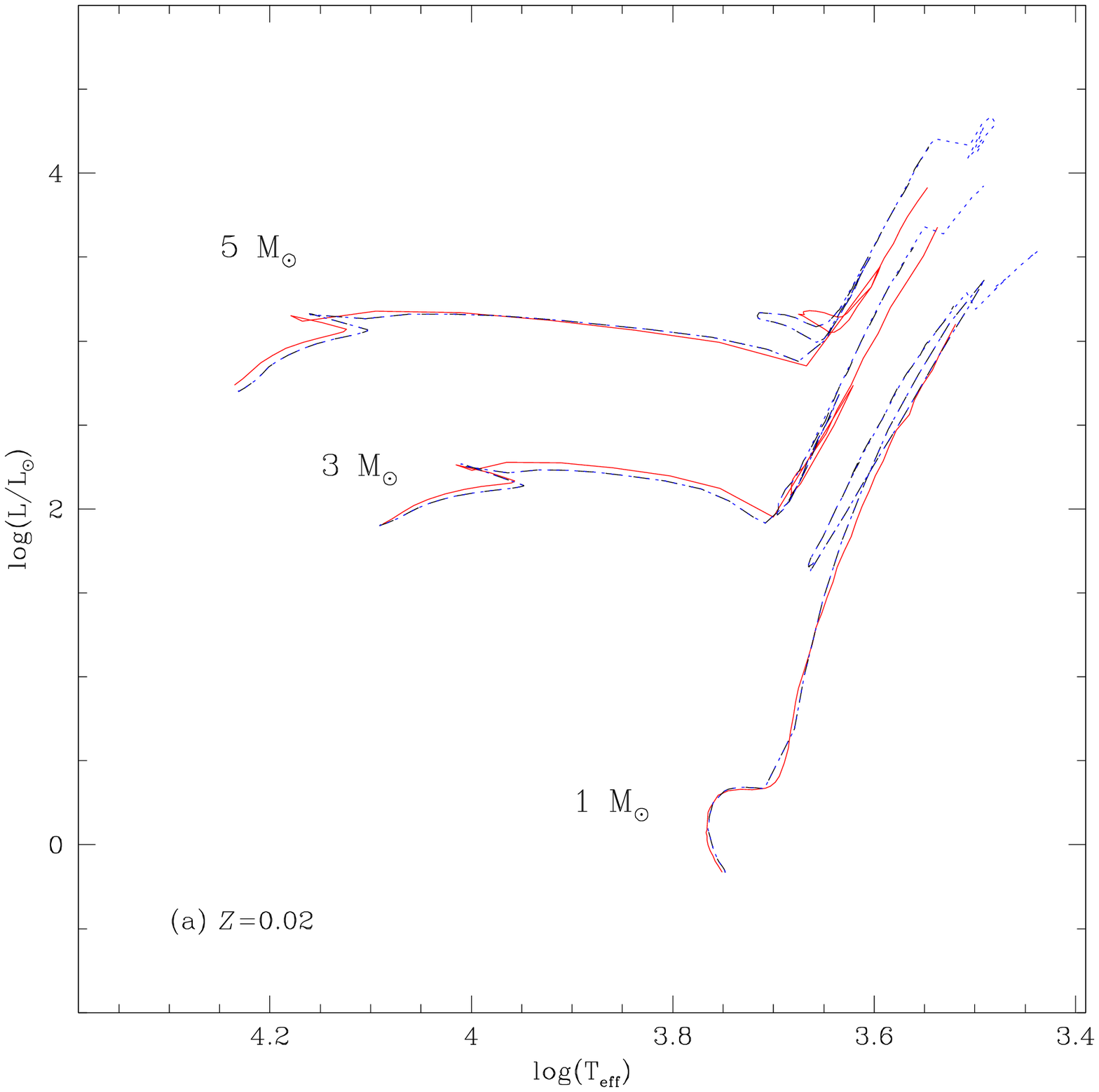}
\caption{Evolutionary tracks for masses of 1, 3, and 5~{\msun} from Geneva (solid), Padova (dashed), and Padova with the TIP-AGB included (dotted). (a) $Z = 0.02$, (b) $Z = 0.008$, and (c) $Z=0.004$.
\label{fig3}}
\end{figure}


\begin{figure}
\plotone{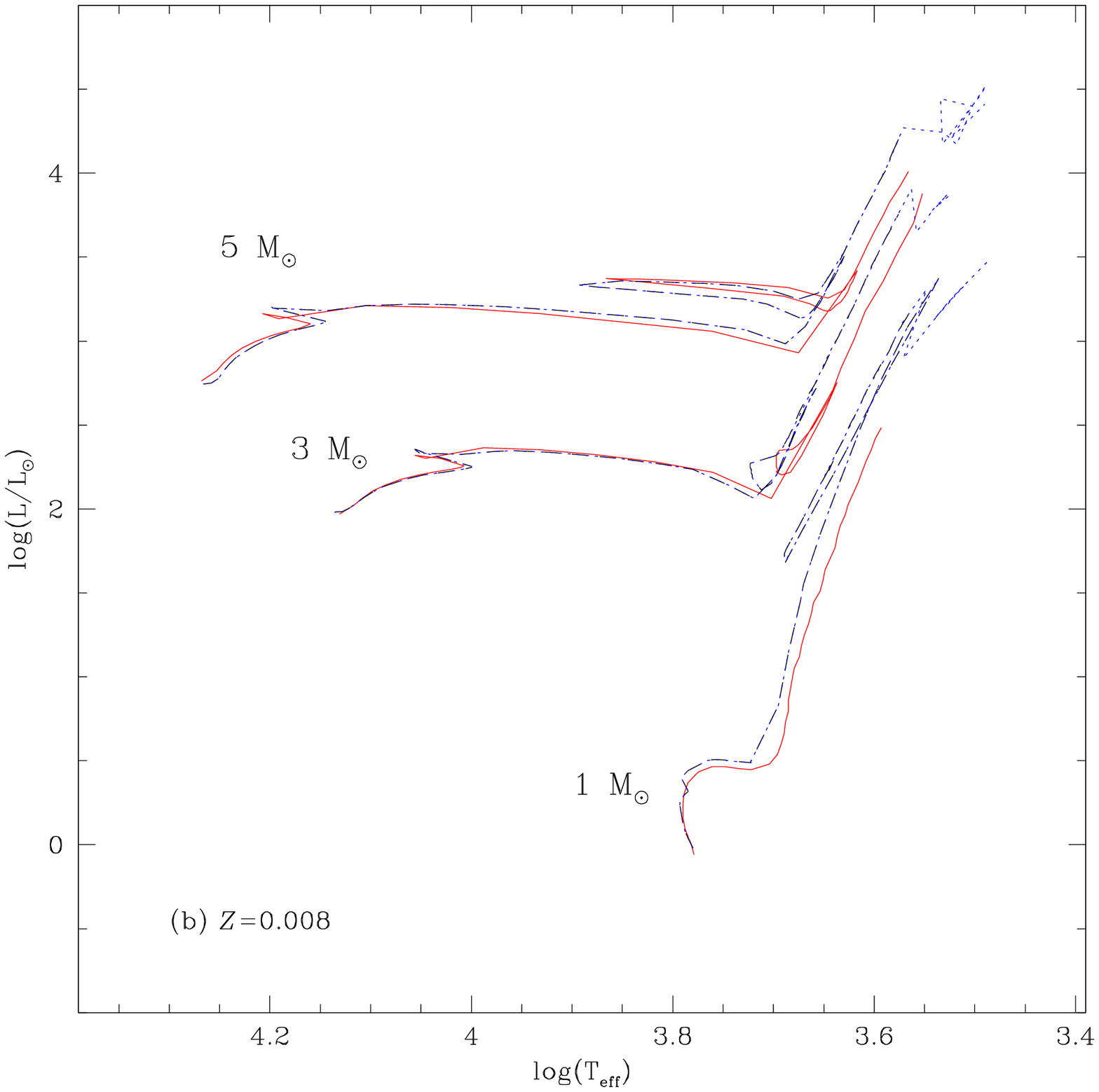}
\end{figure}


\begin{figure}
\plotone{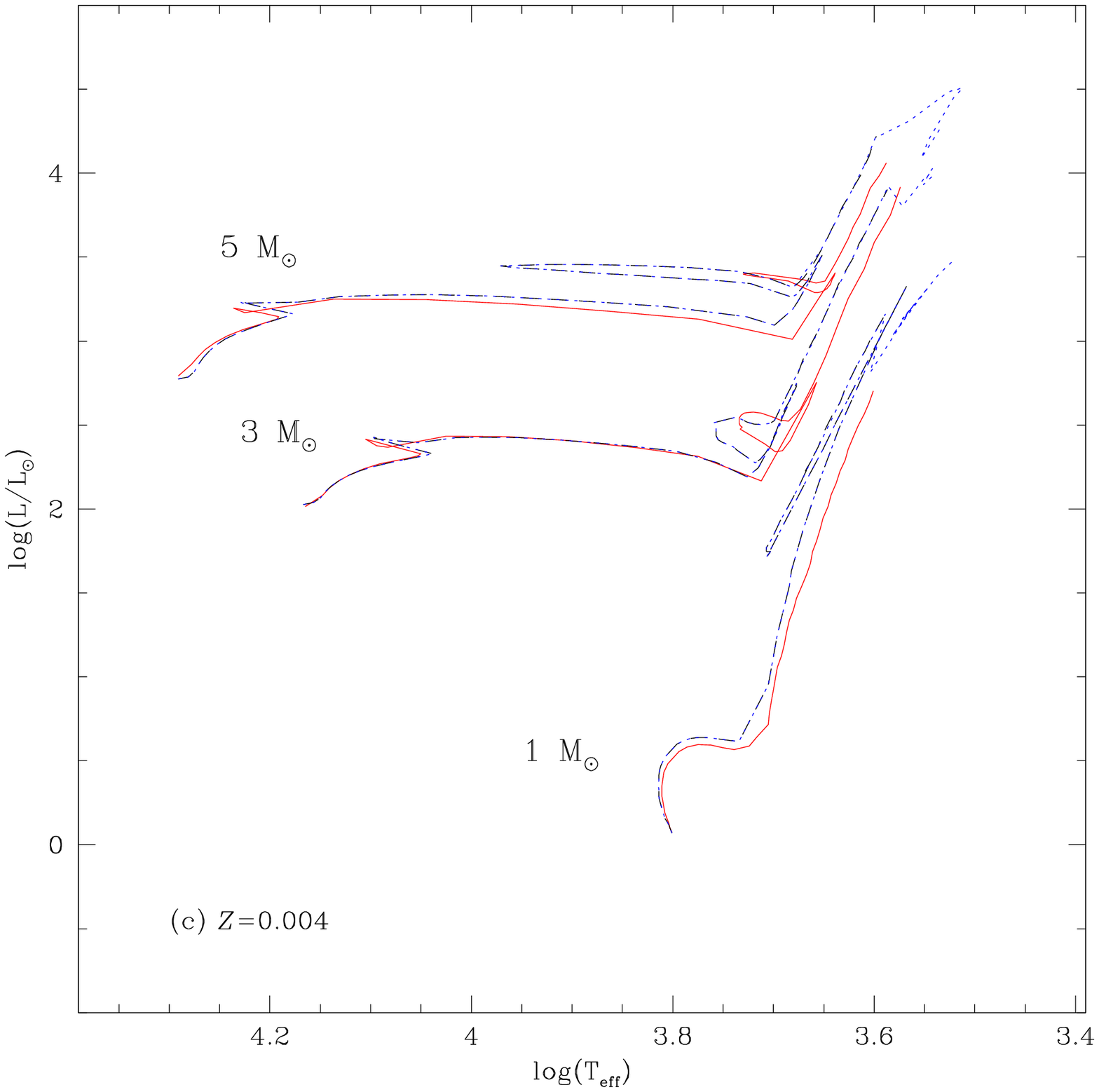}
\end{figure}


\begin{figure}
\plotone{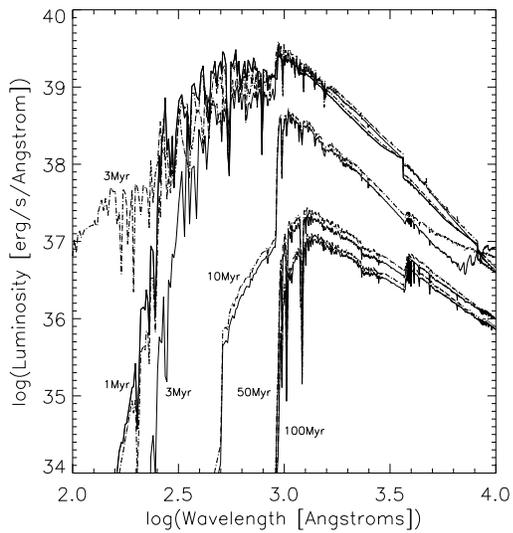}
\caption{Spectral energy distributions for young populations with solar composition. Solid: Geneva; dashed: original Padova; dotted: Padova$+$TIP-AGB. The far-UV continuum produced by the Geneva and Padova models at 3 Myr differs dramatically and has been labeled twice for clarity. Note that the Padova and the Padova$+$TIP-AGB models coincide prior to the occurrence of the first AGB stars, producing a ``dashed-dotted'' graph.
\label{fig4}}
\end{figure}


\begin{figure}
\plotone{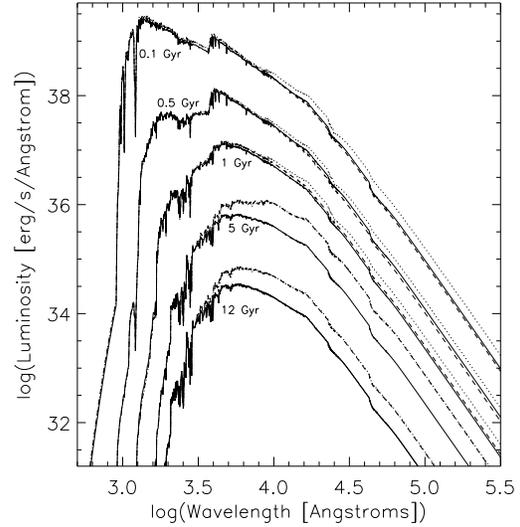}
\caption{Spectral energy distributions for intermediate-age and old populations. Same as for Fig.~4, except for the different age range. The spectra for 0.1, 0.5, 1, and 5 Gyr are displaced in $\log L$ by 2.5, 2.0, 1.5, and 1.0 dex from their original value, respectively. 
\label{fig5}}
\end{figure}


\begin{figure}
\plotone{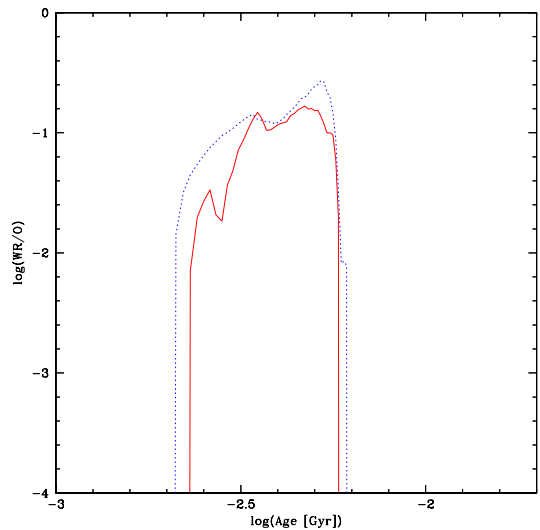}
\caption{Evolution of the W-R/O star ratio. Solid: Geneva; dotted: Padova. Solar composition. 
\label{fig6}}
\end{figure}


\begin{figure}
\plotone{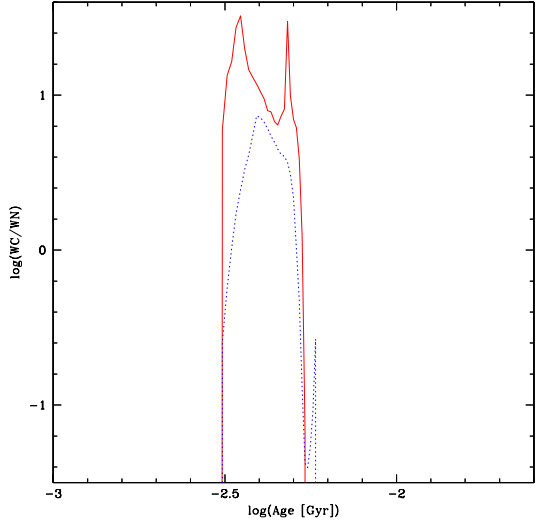}
\caption{Same as Fig.~6, but for the ratio of WC over WN stars. 
\label{fig7}}
\end{figure}


\begin{figure}
\plotone{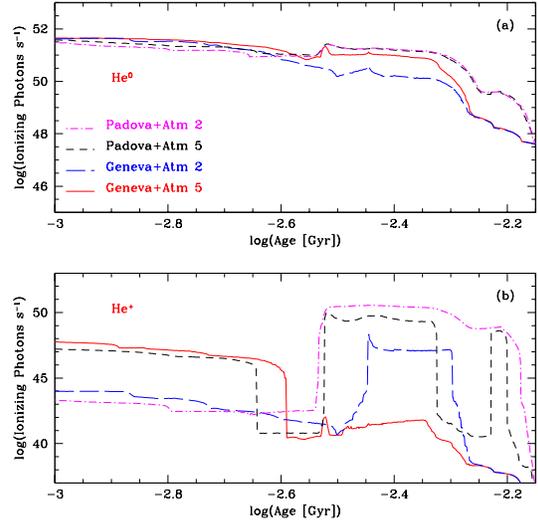}
\caption{(a) Number of photons with wavelength below 504 {\AA} at solar composition. Solid line: Geneva tracks with the extended atmospheres of Smith et al. (2002); long-dashed: Geneva tracks with the plane-parallel atmospheres of Lejeune et al. (1997; 1998); short-dashed: Padova tracks with extended atmospheres; dash-dotted: Padova tracks with plane-parallel atmospheres. W-R stars first appear at age $\sim 3$ Myr. (b) Same as in (a) but for the number of photons with wavelength below 228 {\AA}.
\label{fig8}}
\end{figure}


\begin{figure}
\plotone{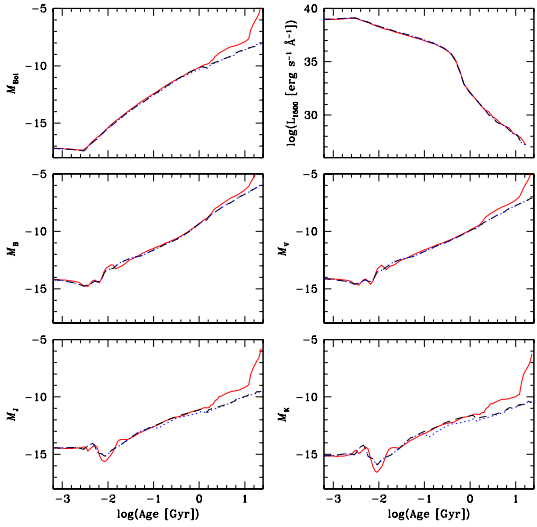}
\caption{$M_{\rm{Bol}}$, $L_{\rm{1500}}$, $M_{\rm{B}}$, $M_{\rm{V}}$, $M_{\rm{J}}$, and $M_{\rm{K}}$ vs. time for models with the Geneva (solid), Padova (dashed), and Padova$+$TIP-AGB (dotted) tracks. Luminosity units are magnitudes, except for $L_{\rm{1500}}$. The contribution from AGB stars becomes noticeable between 0.1 and 3~Gyr at the longest wavelengths. Solar chemical composition. 
\label{fig9}}
\end{figure}


\begin{figure}
\plotone{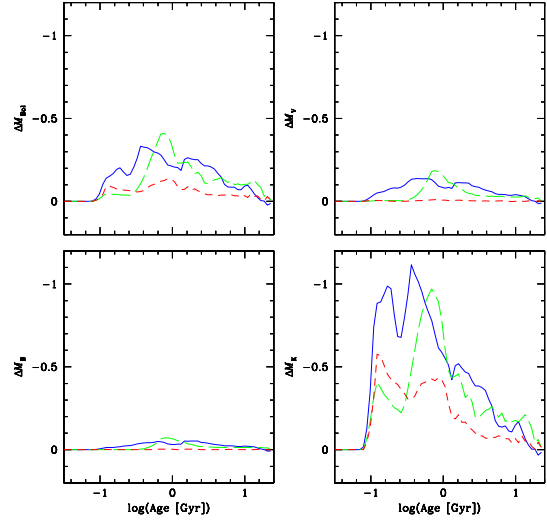}
\caption{Contribution of the TIP-AGB phase in the Padova tracks to the total light in different bands. Solar chemical composition (short dashed line), $Z = 0.008$ (long dashed), and $Z = 0.004$ (solid).
\label{fig10}}
\end{figure}


\begin{figure}
\plotone{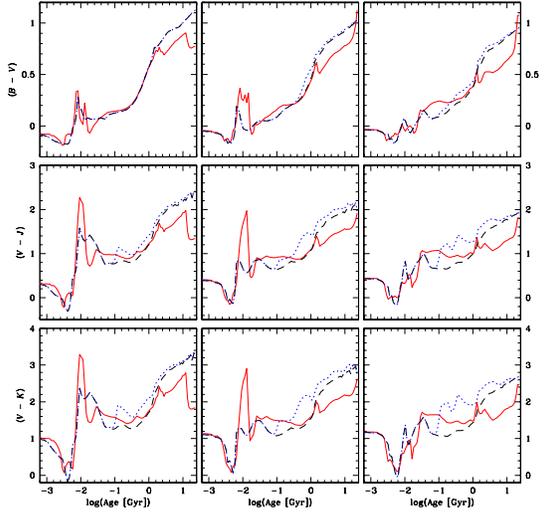}
\caption{Evolution of ($B - V$) (top), ($V - J$) (center), and ($V - K$) (bottom) with time. Left panels: $Z = 0.02$; center panels: $Z = 0.008$; right panels: $Z = 0.004$. Line types as in Fig.~9. 
\label{fig11}}
\end{figure}


\begin{figure}
\plotone{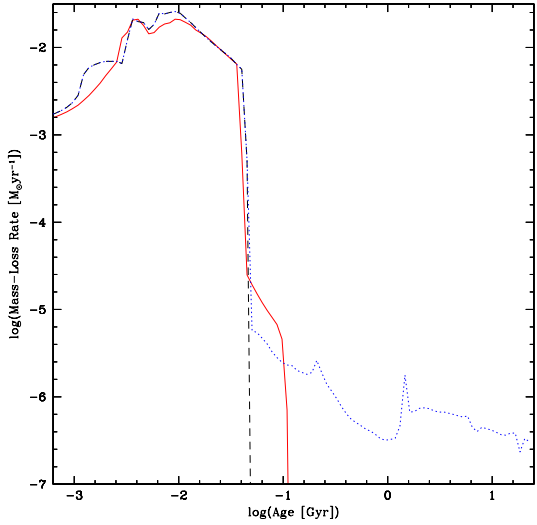}
\caption{Evolution of the mass-loss rate for the three sets of tracks. Notice the extension of $\dot M$ around $\log(Age) > -1.3$ where AGB stars with masses $0.9 < M/M_\sun < 5.0$ have strong mass loss. Same notation as before. Solar chemical composition. 
\label{fig12}}
\end{figure}


\begin{figure}
\plotone{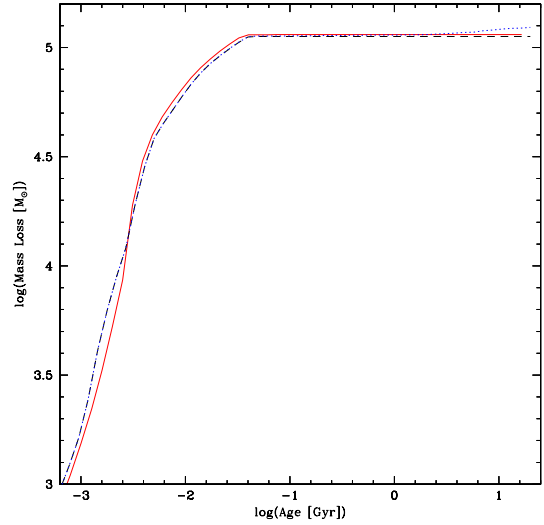}
\caption{Evolution of the total mass loss. Most of the mass loss 
occurs for young populations. Models with no mass loss for the TIP-AGB have a 
constant total mass loss for old populations. AGB stars provide an incremental increase at very old ages. Same notation as before. Solar chemical composition.
\label{fig13}}
\end{figure}


\begin{figure}
\plotone{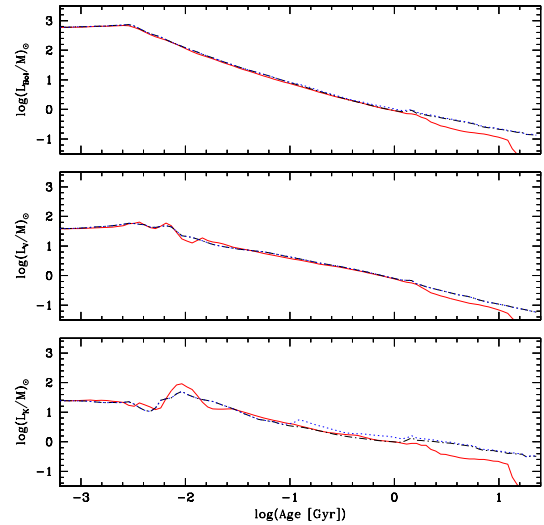}
\caption{The mass-to-light ratio for $L_{\rm{Bol}}$, $L_{\rm{V}}$, and $L_{\rm{K}}$. Same notation as before. Solar chemical composition. 
\label{fig14}}
\end{figure}

\clearpage

\begin{figure}
\plotone{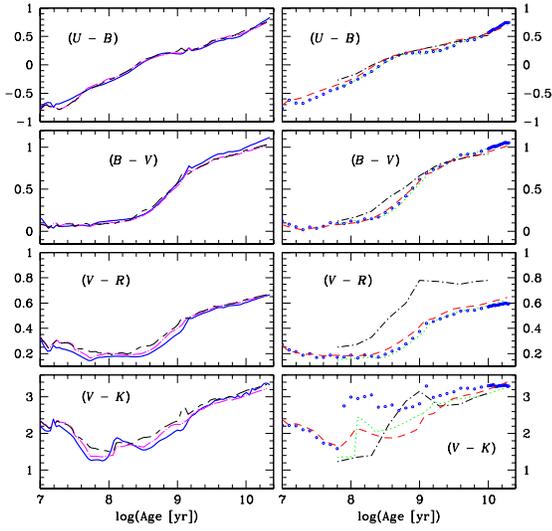}
\caption{Color evolution for different models with solar composition. Bold solid line: this work; long-short-dashed: GALAXEV with Padova (1994) tracks; long-dash-dotted: GALAXEV with Padova (2000) tracks; open small circles: Spectral and Padova (1994) isochrones; dotted: Spectral and Padova (2000) isochrones; short-dash-dotted: Maraston (1998); short-dashed: Fioc \& Rocca-Volmerange (1997). Starburst99 and GALAXEV are plotted together in the left panels because of their almost identical physical ingredients.
\label{fig15}}
\end{figure}


\begin{figure}
\plotone{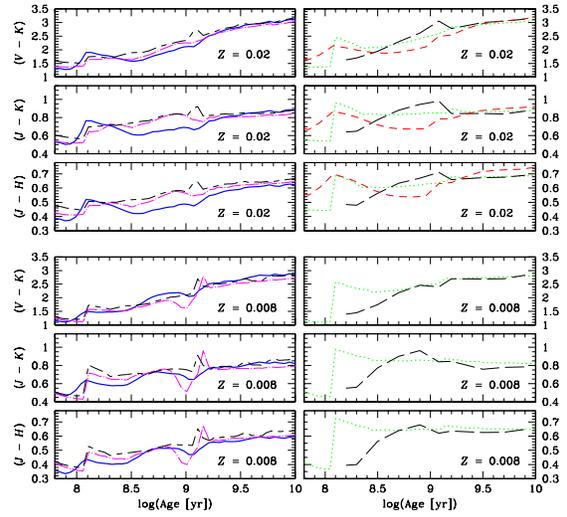}
\caption{Near-IR color evolution for solar composition (upper panels) and $Z = 0.008$ (lower panels). Bold solid line: this work; long-short-dashed: GALAXEV with Padova (1994) tracks; long-dash-dotted: GALAXEV with Padova (2000) tracks; dotted: Spectral and Padova (2000) isochrones; long-dashed: Mouhcine \& Lan\c con (2002); short-dashed: Fioc \& Rocca-Volmerange (1997).
\label{fig16}}
\end{figure}


\begin{figure}
\plotone{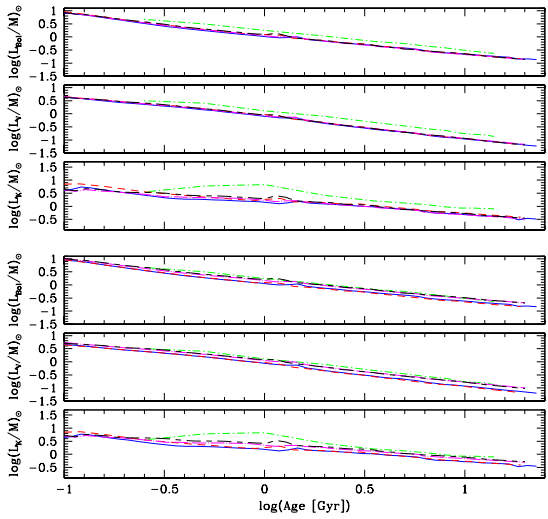}
\caption{Evolution of the mass-to-light ratio including the full AGB phase for $L_{\rm {Bol}}$, $L_{\rm {V}}$ and $L_{\rm {K}}$ at solar composition. Upper set of panels: fixed mass for all models except Maraston (1998); lower set: variable mass for all models except Fioc \& Rocca-Volmerange (1997). Solid line: this work; long-short-dashed: GALAXEV with Padova (1994) tracks; long-dash-dotted: GALAXEV with Padova (2000) tracks; short-dash-dotted: Maraston; short-dashed: Fioc \& Rocca-Volmerange.
\label{fig17}}
\end{figure}


\begin{figure}
\plotone{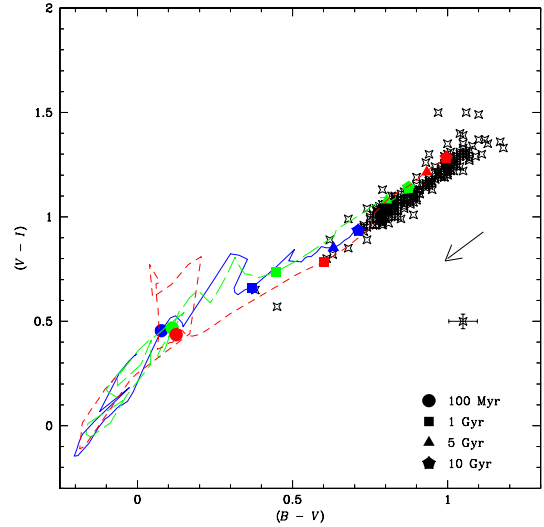}
\caption{($V - I$) vs. ($B - V$) two-color diagram of NGC~5128. The colors predicted by the models are indicated by the three curves for $Z = 0.0004$ (solid), 0.004 (long-dashed), and 0.02 (short-dashed). Tick marks indicate ages of 0.1, 1, 5, and 10~Gyr. The points (together with the schematic error cross at 0,1.5) represent the globular cluster photometry of Peng et al. (2004a). The arrow is the reddening vector for a foreground reddening of $E(B - V) = 0.115$.
\label{fig18}}
\end{figure}


\begin{figure}
\plotone{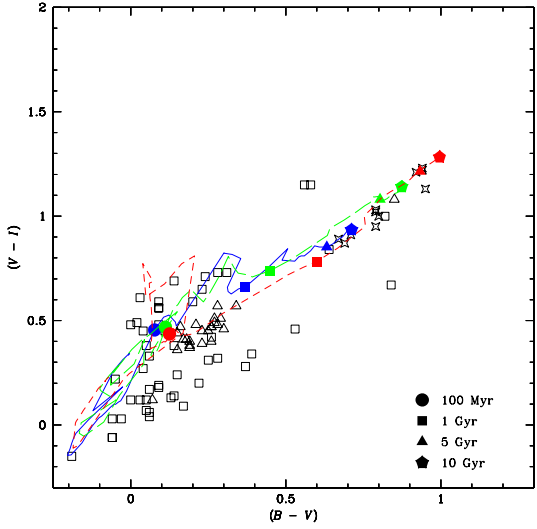}
\caption{Same as Fig.~18, but for the star clusters in NGC~4038/39 of Whitmore et al. (1999). Open stars: old globular clusters; open triangles: intermediate-age clusters; open squares: young clusters. No reddening correction was applied. The models are identical to those in Fig.~18.
\label{fig19}}
\end{figure}


\begin{figure}
\plotone{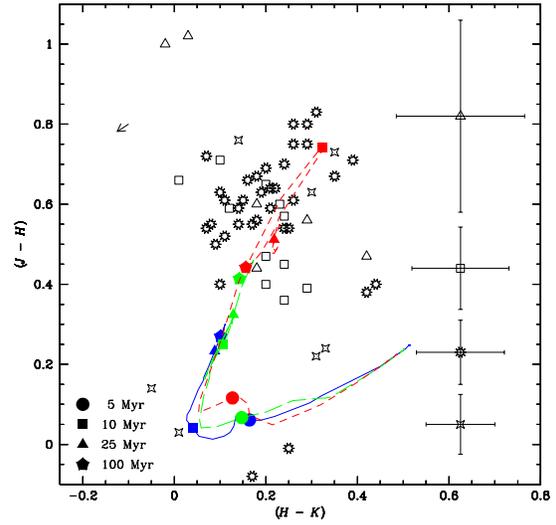}
\caption{Color-color diagram of four samples of BCDs compared with starburst models of $Z = 0.001$ (solid line, blue), 0.004 (long-dashed line, green), and solar $Z = 0.02$ (short-dashed line, red) with Geneva tracks. Filled symbols indicate the ages of the models. Open symbols show the data: triangles (Doublier et al.; 2001), squares (Noeske et al.; 2003), starred circles (Thuan; 1983), and stars (Telles; 2004). The error bars are the dispersion value in each sample. The solid vector indicates the reddening correction for $A_V = 0.25$
\label{fig20}}
\end{figure}


\begin{figure}
\plotone{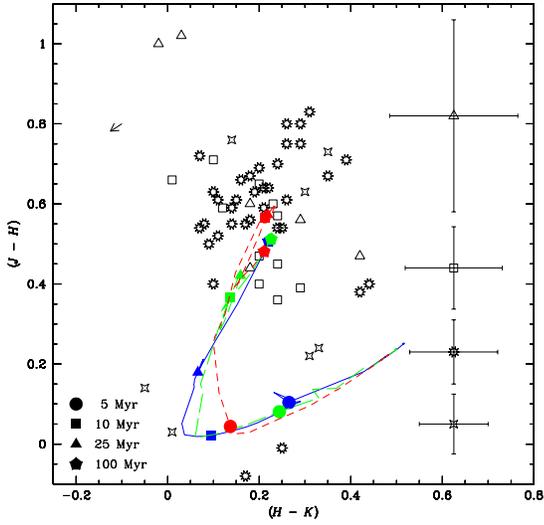}
\caption{Same as Fig.~20 for starburst models of $Z = 0.0004$ (solid line, 
blue), 0.004 (long-dashed line, green), and $Z = 0.02$ (short-dashed line, 
red) with Padova tracks.
\label{fig21}}
\end{figure}


\begin{figure}
\plotone{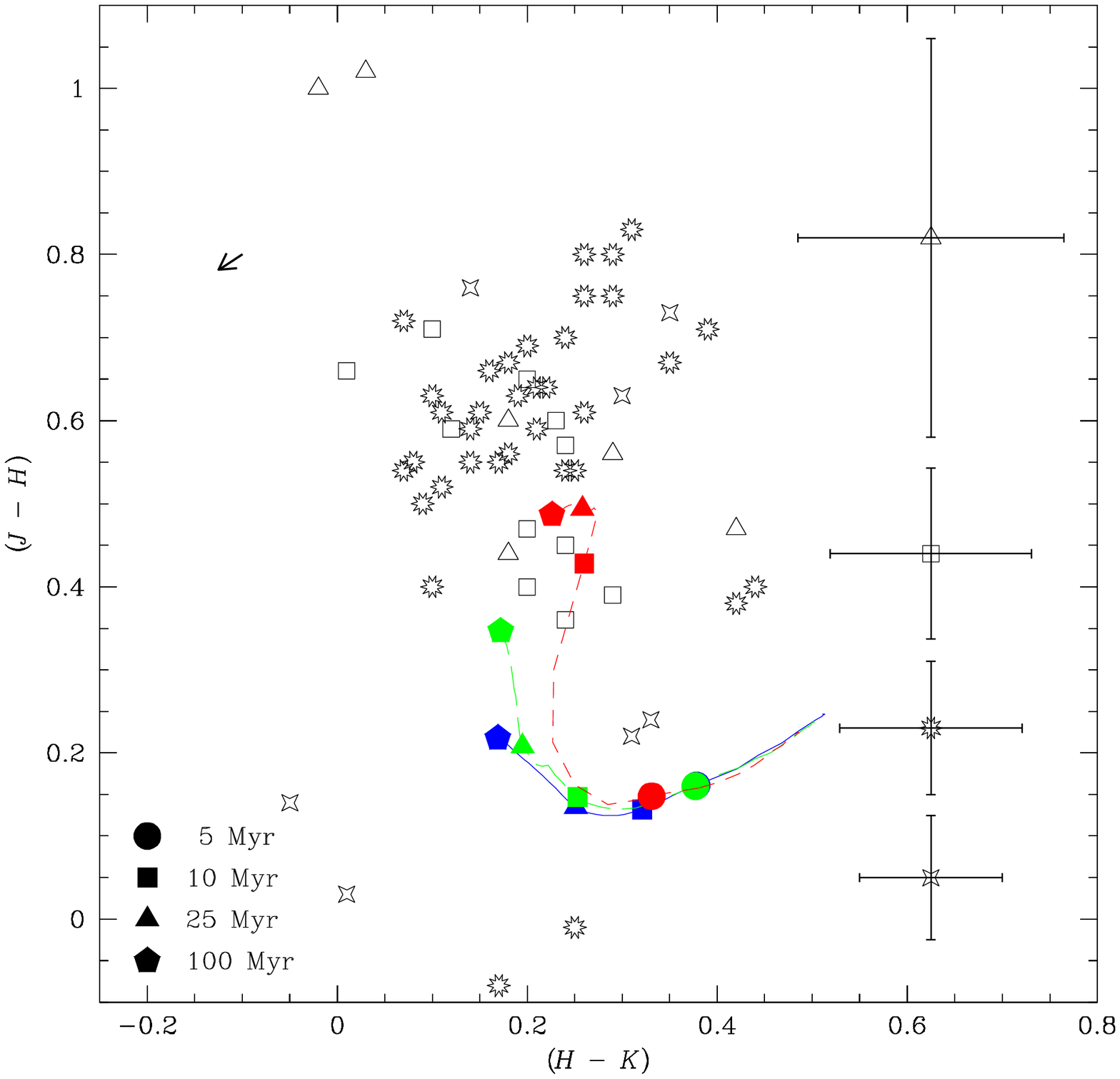}
\caption{Same as Fig.~20 for continuous star formation models of $Z = 0.001$ 
(solid line, blue), 0.004 (long-dashed line, green), and $Z = 0.02$ 
(short-dashed line, red) with Geneva tracks.
\label{fig22}}
\end{figure}


\begin{figure}
\plotone{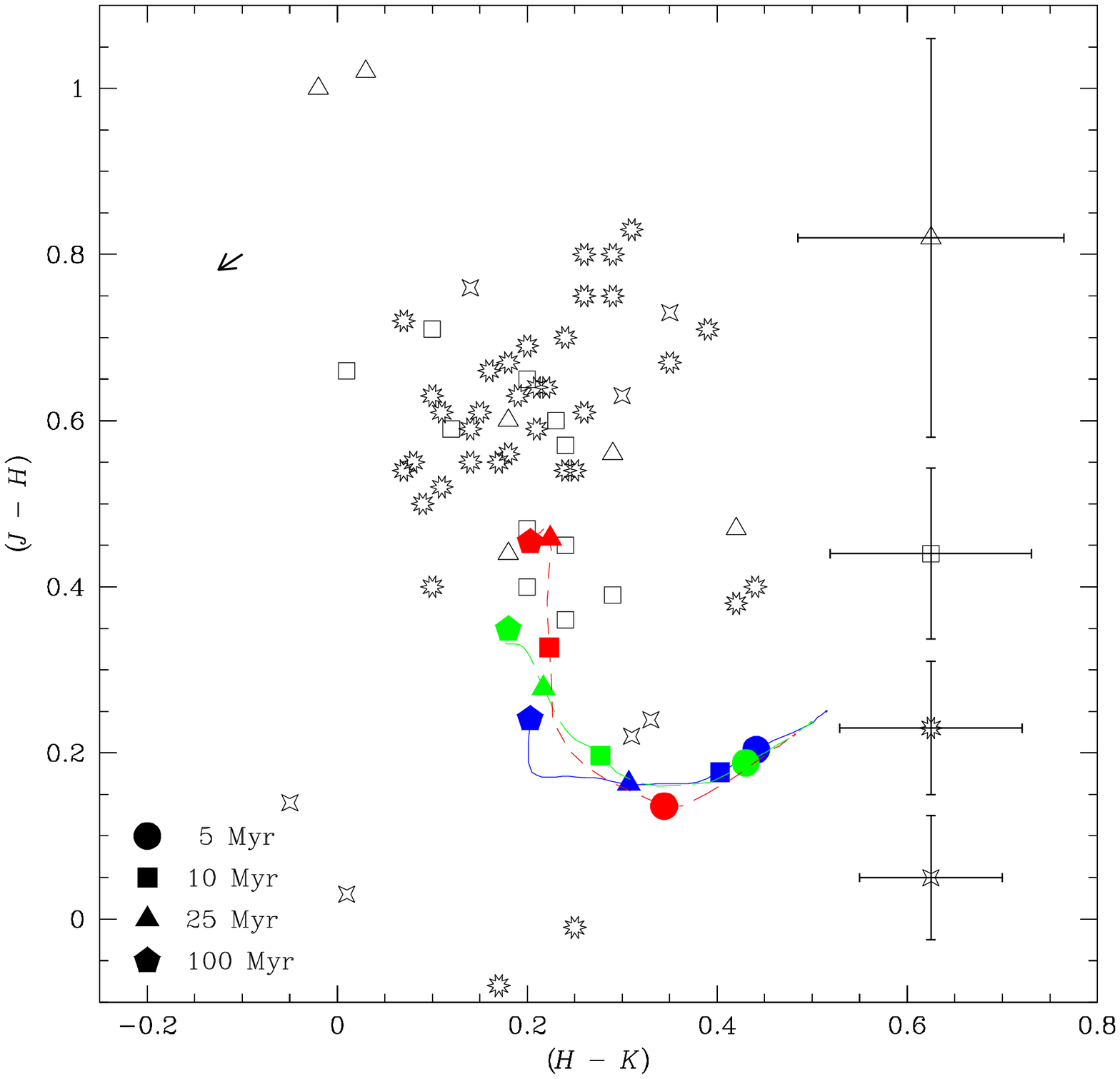}
\caption{Same as Fig.~20 for continuous star formation models of $Z = 0.0004$ 
(solid line, blue), 0.004 (long-dashed line, green), and $Z = 0.02$ 
(short-dashed line, red) with Padova tracks.
\label{fig23}}
\end{figure}

\clearpage

\clearpage

\begin{deluxetable}{ccc}
\tabletypesize{\scriptsize}
\tablecaption{Adopted solar abundances of the main elements in the Geneva and Padova tracks. 
\label{tbl-1}}
\tablewidth{0pt}

\tablehead{

\colhead{Atomic Number}   &
\colhead{Element}   &  
\colhead{Mass Fraction}   \\
}

\startdata

 1 & H    & $7.06 \times 10^{-1}$ \\
 2 & He   & $2.75 \times 10^{-1}$ \\
 3 & Li   & $9.35 \times 10^{-9}$ \\
 4 & Be   & $1.66 \times 10^{-10}$ \\
 5 & B    & $4.73 \times 10^{-9}$ \\
 6 & C    & $3.03 \times 10^{-3}$ \\
 7 & N    & $1.11 \times 10^{-3}$ \\
 8 & O    & $9.59 \times 10^{-3}$ \\
 9 & F    & $4.05 \times 10^{-7}$ \\
10 & Ne   & $1.62 \times 10^{-3}$ \\
11 & Na   & $3.34 \times 10^{-5}$ \\
12 & Mg   & $5.15 \times 10^{-4}$ \\
13 & Al   & $5.81 \times 10^{-5}$ \\
14 & Si   & $6.53 \times 10^{-4}$ \\
16 & S    & $3.96 \times 10^{-4}$ \\
25 & Mn   & $1.33 \times 10^{-5}$ \\
26 & Fe   & $1.17 \times 10^{-3}$ \\

\enddata

\end{deluxetable}

\clearpage

\begin{deluxetable}{lccccc}
\tabletypesize{\scriptsize}
\tablecaption{Mass fractions of H, He, C, N, and O used in the Geneva and Padova tracks. 
\label{tbl-2}}
\tablewidth{0pt}

\tablehead{

\multicolumn{6}{c}{Padova Tracks} \\

\colhead{Element}  & 
\multicolumn{1}{c}{$Z=0.0004$} & 
\multicolumn{1}{c}{$Z=0.004$} & 
\multicolumn{1}{c}{$Z=0.008$} & 
\multicolumn{1}{c}{$Z=0.02$} & 
\multicolumn{1}{c}{$Z=0.05$}  \\}
\startdata
 $^1$H    & 7.70(-01) & 7.56(-01) & 7.42(-01) & 7.00(-01) & 5.98(-01) \\
 $^2$He   & 2.30(-01) & 2.40(-01) & 2.50(-01) & 2.80(-01) & 3.52(-01) \\
 $^6$C    & 9.90(-05) & 9.89(-04) & 1.98(-03) & 4.94(-03) & 1.24(-02) \\
 $^7$N    & 2.50(-05) & 2.48(-04) & 4.97(-04) & 1.24(-03) & 3.10(-03) \\
 $^8$O    & 2.11(-04) & 2.11(-03) & 4.23(-03) & 1.06(-02) & 2.64(-02) \\

\tableline

\multicolumn{6}{c}{Geneva Tracks} \\

\colhead{}  & 
\multicolumn{1}{c}{$Z=0.001$} & 
\multicolumn{1}{c}{$Z=0.004$} & 
\multicolumn{1}{c}{$Z=0.008$} & 
\multicolumn{1}{c}{$Z=0.02$} & 
\multicolumn{1}{c}{$Z=0.04$}  \\

\tableline

 $^1$H    & 7.56(-01) & 7.44(-01) & 7.28(-01) & 6.80(-01) & 6.20(-01) \\
 $^2$He   & 2.43(-01) & 2.52(-01) & 2.64(-01) & 3.00(-01) & 3.40(-01) \\
 $^6$C    & 2.43(-04) & 9.73(-04) & 1.95(-03) & 4.87(-03) & 9.73(-03) \\
 $^7$N    & 6.20(-05) & 2.47(-04) & 4.95(-04) & 1.24(-03) & 2.47(-03) \\
 $^8$O    & 5.27(-04) & 2.11(-03) & 4.22(-03) & 1.05(-02) & 2.11(-02) \\

\enddata

\end{deluxetable}

\clearpage

\begin{table}
\caption{Properties of the two sets of tracks.} 
\label{tbl-3}
\begin{center}\scriptsize
\begin{tabular}{lcc}
Property & Geneva & Padova$+$TIP-AGB \\
\hline
\hline
$Z/Z_\odot$ & 0.05, 0.2, 0.4, 1, 2 & 0.02, 0.2, 0.4, 1, 2.5 \\
Upper Mass & 120 & 120 \\
Lower Mass & 0.8 & 0.15 \\
Grid Points & 51 & 84 \\
Phases & 
MS, RGB, ZAHB\tablenotemark{a}, and TP-AGB\tablenotemark{b} & 
MS, RGB, HB, TP-AGB, and TIP-AGB\tablenotemark{c} \\
\hline
\hline
\end{tabular}

\tablenotetext{a}{The HB phase for intermediate-mass stars in Geneva models included only at the beginning of the helium burning on the zero-age horizontal branch (ZAHB).}
\tablenotetext{b}{For stars with $M < 1$~{\msun}, the tracks have the RGB phase for $Z = 0.004$ and 0.008. Otherwise the evolution is limited to the early evolution after the MS. The RGB is the last phase modeled in stars with $1 \le M/M_\odot < 1.7$.}
\tablenotetext{c}{For stars with $0.6 \le {\rm M}/{\rm{M}}_\odot \le 0.9$, the tracks follow the evolution up to the RGB phase only.}


\end{center}
\end{table}

\clearpage

\begin{deluxetable}{lcccccc}
\tabletypesize{\tiny}
\rotate
\tablecaption{AGB implementation in Starburst99 and other evolutionary synthesis codes. 
\label{tbl-4}}
\tablewidth{0pt}
\tablehead{
\colhead{Designation} & 
\colhead{Stellar models}   & 
\colhead{AGB phase} &
\colhead{$M_{\rm {max}}$} &
\colhead{SEDs}  &
\colhead{Photometry} &
\colhead{Reference} \\
}

\startdata

Starburst99 & 
Padova (2000) tracks & 
Vassiliadis \& Wood (1993) & 
5 &
Lejeune et al. (1997;1998) &
Johnson-Cousins &
This work \\
GALAXEV (2003) &
Padova (2000) tracks &
Vassiliadis \& Wood (1993) &
5 &
Le Borgne et al. (2003) &
Johnson-Cousins &
Bruzual \& Charlot (2003) \\
GALAXEV (2003) &
Padova (1994) tracks &
Vassiliadis \& Wood (1993) &
5 &
Le Borgne et al. (2003) &
Johnson-Cousins &
Bruzual \& Charlot (2003) \\
Spectral &
Padova (2000) isochrones &
Groenewegen \& de Jong (1993)\tablenotemark{a} &
5 &
Lejeune et al. (1997; 1998) &
Johnson-Cousins &
V\'azquez et al. (2003) \\
Spectral &
Padova (1994) isochrones &
Boothroyd \& Sackmann (1988)\tablenotemark{b} &
5 &
Lejeune et al. (1997; 1998) &
Johnson-Cousins &
V\'azquez et al. (2003) \\
Mouhcine \& Lan\c con &
Padova (1994) tracks &
Wagenhuber \& Groenewegen (1998)\tablenotemark{c} &
7 &
Lan\c con \& Wood (2000) &
Johnson-Cousins &
Mouhcine \& Lan\c con (2002) \\
Maraston &
Castellani et al. (1992) &
Bl\"ocker \& Sch\"onberner (1991) &
8 &
Kurucz (1979) &
Johnson &
Maraston (1998) \\
PEGASE &
Padova (1994) tracks &
Sch\"onberner (1983), Bl\"ocker (1995) &
6 &
Gunn \& Stryker (1983) &
Johnson-Cousins &
Fioc \& Rocca-Volmerange (1997) \\

\enddata

\tablenotetext{a}{The mass loss prescription of Vassiliadis \& Wood (1993) is used.}
\tablenotetext{b}{The Reimers (1975) formula for the mass-loss rate is used.}
\tablenotetext{c}{The Bl\"ocker (1995) formula for the mass loss rate is used.}

\end{deluxetable}

\clearpage

\begin{deluxetable}{llcccc}
\tabletypesize{\scriptsize}
\tablecaption{Sample of BCD galaxies used for comparison with our models. 
\label{tbl-5}}
\tablewidth{0pt}
\tablehead{
\colhead{Id} & 
\colhead{Object}   & 
\colhead{$(J - H)$} &
\colhead{$\Delta (J - H)$} &
\colhead{$(H - K)$}  &
\colhead{$\Delta (H - K)$} \\
}

\startdata

 1KN  & Tol~3            &  0.60 & 0.07 &  0.23 & 0.06 \\
 2KN & Tol~1400-411c    &  0.36 & 0.08 &  0.24 & 0.09 \\
 3KN & Pox~4c           &  0.39 & 0.13 &  0.29 & 0.13 \\
 4KN & Pox~4B           &  0.40 & 0.15 &  0.20 & 0.15 \\
 5KN & UM~448           &  0.65 & 0.10 &  0.20 & 0.11 \\
 6KN & UM~461           &  0.47 & 0.06 &  0.20 & 0.06 \\
 7KN & He~2-10          &  0.57 & 0.06 &  0.24 & 0.07 \\
 8KN & IC~4662c         &  0.59 & 0.12 &  0.12 & 0.12 \\
 9KN & Mrk~178          &  0.45 & 0.15 &  0.24 & 0.16 \\
10KN & Mrk~1329         &  0.66 & 0.09 &  0.01 & 0.10 \\
11KN & Haro~14          &  0.71 & 0.12 &  0.10 & 0.12 \\
\hline
\hline
12VD  & Haro~14          &  0.60 & 0.24 &  0.18 & 0.14 \\
13VD  & Mrk~996          &  0.47 & 0.24 &  0.42 & 0.14 \\
14VD  & Tol~0610-387     &  0.44 & 0.24 &  0.18 & 0.14 \\
15VD  & Tol~0957-279     &  1.02 & 0.24 &  0.03 & 0.14 \\
16VD  & Fairall~301      &  0.56 & 0.24 &  0.29 & 0.14 \\
17VD  & Tol~0954-293     &  1.00 & 0.24 & -0.02 & 0.14 \\
\hline
\hline
18TT  & Mrk~36           &  0.38 & 0.14 &  0.42 & 0.15 \\ 
19TT  & Mrk~49           &  0.54 & 0.14 &  0.25 & 0.15 \\
20TT  & Mrk~59           &  0.40 & 0.14 &  0.44 & 0.15 \\
21TT  & Mrk~86           &  0.69 & 0.03 &  0.20 & 0.04 \\
22TT  & Mrk~89           &  0.72 & 0.06 &  0.07 & 0.07 \\
23TT  & Mrk~116          &  0.52 & 0.18 &  0.11 & 0.27 \\
24TT  & Mrk~140          &  0.70 & 0.04 &  0.24 & 0.06 \\
25TT  & Mrk~169          &  0.67 & 0.03 &  0.35 & 0.04 \\
26TT  & Mrk~209          &  0.40 & 0.08 &  0.10 & 0.12 \\
27TT  & Mrk~297          &  0.75 & 0.04 &  0.29 & 0.06 \\
28TT  & Mrk~313          &  0.80 & 0.04 &  0.26 & 0.06 \\
29TT  & Mrk~314          &  0.50 & 0.08 &  0.09 & 0.08 \\
30TT  & Mrk~324          &  0.63 & 0.05 &  0.19 & 0.07 \\
31TT  & Mrk~328          &  0.64 & 0.06 &  0.21 & 0.07 \\
32TT  & Mrk~370          &  0.61 & 0.04 &  0.26 & 0.06 \\
33TT  & Mrk~401          &  0.75 & 0.03 &  0.26 & 0.04 \\
34TT  & Mrk~527          &  0.83 & 0.04 &  0.31 & 0.06 \\
35TT  & Mrk~600          &  0.54 & 0.11 &  0.07 & 0.11 \\
36TT  & I-Zw~123         &  0.54 & 0.15 &  0.24 & 0.15 \\
37TT  & II-Zw~70         &  0.55 & 0.04 &  0.17 & 0.05 \\
38TT  & II-Zw~71         &  0.61 & 0.04 &  0.15 & 0.06 \\
39TT  & III-Zw~102       &  0.80 & 0.04 &  0.29 & 0.06 \\
40TT  & Haro~2           & -0.01 & 0.07 &  0.25 & 0.04 \\
41TT  & Haro~3           & -0.08 & 0.04 &  0.17 & 0.04 \\
42TT  & Haro~9           &  0.59 & 0.14 &  0.14 & 0.14 \\
43TT  & Haro~14          &  0.61 & 0.05 &  0.11 & 0.06 \\
44TT  & Haro~15          &  0.59 & 0.04 &  0.21 & 0.06 \\
45TT  & Haro~18          &  0.63 & 0.05 &  0.10 & 0.07 \\
46TT  & Haro~20          &  0.64 & 0.07 &  0.22 & 0.07 \\
47TT  & Haro~21          &  0.66 & 0.05 &  0.16 & 0.06 \\
48TT  & Haro~23          &  0.56 & 0.04 &  0.18 & 0.08 \\
49TT  & Haro~28          &  0.67 & 0.14 &  0.18 & 0.14 \\
50TT  & Haro~36          &  0.55 & 0.15 &  0.08 & 0.17 \\
51TT  & Haro~38          &  0.55 & 0.15 &  0.14 & 0.15 \\
52TT  & Haro~43          &  0.71 & 0.30 &  0.39 & 0.29 \\
\hline
\hline
53ET  & C0828a           &  0.73 & 0.05 &  0.35 & 0.05 \\
54ET  & II-Zw~40         &  0.76 & 0.05 &  0.14 & 0.05 \\
55ET  & I-Zw~18          &  0.24 & 0.05 &  0.33 & 0.05 \\
56ET  & Mrk~36           &  0.22 & 0.05 &  0.31 & 0.05 \\
57ET  & UM~448           &  0.63 & 0.10 &  0.30 & 0.10 \\
58ET  & UM~461           &  0.03 & 0.10 &  0.01 & 0.10 \\
59ET  & UM~559           &  0.14 & 0.10 & -0.05 & 0.10 \\

\enddata

\end{deluxetable}






\end{document}